%% file: main.tex
\documentclass[11pt]{article}
\usepackage[letterpaper,margin=1.1in]{geometry}
\pdfoutput=1
\usepackage[square,sort,comma,numbers]{natbib}
\usepackage{amssymb}
\usepackage{graphicx}
\usepackage{sidecap}

\usepackage{url}
\usepackage{graphicx}
\usepackage{times}
\usepackage{setspace}
\usepackage{latexsym}
\usepackage{algorithm}
\usepackage[noend]{algpseudocode}
\usepackage{wrapfig}
\usepackage{sidecap}
\usepackage{booktabs}
\usepackage{aas_macro_long}
\usepackage{caption}
\usepackage{subcaption}
\usepackage[inline]{enumitem}
\usepackage{amsmath}

\def\clq{\mbox{\sc Clique}}

\def\clqh{\mbox{\sc CliqueHeu}}

\def\rowspace{-1pt}

\def\algospacing{1.0}

\begin{document}

\title{Fast Algorithms for the Maximum Clique Problem on Massive Graphs with Applications
to Overlapping Community Detection}

\author{Bharath Pattabiraman$^{\rm \ast \ddag}$, Md. Mostofa Ali Patwary$^{\rm \ast}$, 
Assefaw H. Gebremedhin$^{\rm \dag \ddag}$, \\
Wei-keng Liao$^{\rm \ast}$, and Alok Choudhary\footnote{Northwestern University, Evanston, IL 60208.
$^{\rm \dag}$Washington State University, Pullman, WA 99163.
$^{\rm \ddag}$Corresponding authors: bpa342@eecs.northwestern.edu, 
assefaw@eecs.wsu.edu.}
}







\maketitle

\date{}

\begin{abstract}

\input{abs}

\end{abstract}


\input{introduction}

\input{relatedwork}

\input{algorithms}

\input{experiments}

\input{parallelization}

\input{applications2}
\input{conclusion}

\section*{Acknowledgements}
This work is supported in part by the following grants: NSF awards CCF-0833131, CNS-0830927, IIS-0905205, CCF-0938000, CCF-1029166, and OCI-1144061; DOE awards DE-FG02-08ER25848, DE-SC0001283, DE-SC0005309, DE-SC0005340, and DE-SC0007456; AFOSR award FA9550-12-1-0458. 
The work of Assefaw Gebremedhin is supported in part by the NSF
award CCF-1218916 and by the DOE award DE-SC0010205.



\newpage
\input{appendix_3}
\input{appendix_2}

\input{appendix_1}

\end{document}

%% file: abs.tex
The maximum clique problem is a well known NP-Hard problem with
applications in data mining, network analysis, information retrieval and many other
areas related to the World Wide Web.
There exist several algorithms for the problem with acceptable runtimes for
certain classes of graphs, but many of them are infeasible for massive graphs. 
We present a new exact algorithm that employs novel pruning techniques and 
is able to find maximum cliques in very large, sparse graphs quickly. 
Extensive experiments on different kinds of synthetic and 
real-world graphs show that our new algorithm can be orders of magnitude 
faster than existing algorithms.
We also present a heuristic that runs orders of magnitude faster than 
the exact algorithm while providing optimal or near-optimal solutions.
We illustrate a simple application of the algorithms in developing methods for
detection of overlapping communities in networks.

%% file: introduction.tex
\section{Introduction}
\label{sec:intro}

A clique in an undirected graph is a subset of vertices in which every two vertices
are adjacent to each other. The {\em maximum} clique problem seeks to find 
a clique of the largest possible size in a given graph.

The maximum clique problem, and the related {\em maximal} clique and 
clique {\em enumeration} problems, find applications in a wide variety of domains,
many intimately related to the World Wide Web. 
A few examples include:  
information retrieval \cite{Augustson:1970:AGT:321607.321608}, 
community detection in networks \cite{Fortunato_2010,cite-key,5586496}, 
spatial data mining \cite{wang2009order},
data mining in bioinformatics \cite{19566964},
disease classification based on symptom correlation \cite{Bonner:1964:CT:1662386.1662389}, 
pattern recognition \cite{1211348},
analysis of financial networks \cite{RePEc:eee:csdana:v:48:y:2005:i:2:p:431-443},
computer vision \cite{Horaud:1989:SCT:68871.68875}, and
coding theory \cite{brouwer}.
More examples of application areas can be found in 
\cite{Gutin2004,citeulike:4058448}.

To get a sense for how clique computation arises in the aforementioned contexts, 
consider a generic data mining or information retrieval problem. A typical objective
here is to retrieve data that are considered similar based on some metric. 
Constructing a graph in which vertices correspond to data items and 
edges connect similar items, a clique in the graph would then give a cluster of similar data. 

The maximum clique problem is NP-Hard \cite{Garey:1979:CIG:578533}.
Most exact algorithms for solving it employ some form of {\it branch-and-bound} approach. 
While branching systematically searches for all candidate solutions, bounding (also known as {\em pruning}) discards fruitless candidates based on a previously computed bound. 
The algorithm of Carraghan and Pardalos \cite{pardalos} is an early example of 
a simple and  effective branch-and-bound algorithm for the maximum clique problem.
More recently, \"{O}sterg{\aa}rd \cite{ostergard} introduced 
an improved algorithm and demonstrated its relative advantages via computational experiments. 
Tomita and Seki \cite{citeulike:7905505}, and later, Konc and Jane\v{z}i\v{c} \cite{konc2007improved}
use upper bounds computed using vertex coloring to enhance the branch-and-bound approach. 
Other examples of branch-and-bound algorithms for the clique problem include 
the works of Bomze et al~\cite{Bomze99themaximum}, Segundo et al.~\cite{SanSegundo}
and Babel and Tinhofer~\cite{babel1990branch}.
Prosser~\cite{prosser2012} in a recent work compares various exact algorithms 
for the maximum clique problem. 


In this paper, we present a new exact branch-and-bound
algorithm for the maximum clique problem that employs 
several new pruning strategies in addition to those used in \cite{pardalos},
\cite{ostergard}, \cite{citeulike:7905505} and \cite{konc2007improved},
making it suitable for massive graphs.
We run our algorithms on a large variety of test graphs and compare its performance with the
algorithms by Carraghan and Pardalos \cite{pardalos}, \"{O}sterg{\aa}rd \cite{ostergard}, Tomita et al. \cite{citeulike:7905505}\cite{walcom}, Konc and Jane\v{z}i\v{c} \cite{konc2007improved}, and Segundo et al.~\cite{SanSegundo}. 
We find our new exact algorithm to be up to orders of magnitude faster on large,
sparse graphs and of comparable runtime on denser graphs.
We also present a hew heuristic, which runs several orders of magnitude faster than the exact algorithm while providing solutions that are optimal or near-optimal for most cases. The algorithms are presented in detail in Section~\ref{sec:algorithms} and the experimental evaluations and comparisons are presented in Section~\ref{sec:experiments}. 

Both the exact algorithm and the heuristic are well-suited for parallelization.
We discuss a simple shared-memory parallelization and present performance 
results showing its promise in Section~\ref{sec:parallelization}. 
We also include (in Section~\ref{sec:applications})
an illustration on how the algorithms can be used as parts of a method
for detecting overlapping communities in networks.
We have made our implementations publicly available\footnote{\url{http://cucis.ece.northwestern.edu/projects/MAXCLIQUE/}}.




%% file: relatedwork.tex
\section{Related Previous Algorithms}
\label{sec:relatedwork}

Given a simple undirected graph $G$, the maximum clique can clearly be obtained by enumerating 
{\em all} of the cliques present in it and picking the largest of them.
Carraghan and Pardalos \cite{pardalos} introduced a simple-to-implement
algorithm that avoids enumerating all cliques and instead
works with a significantly reduced partial enumeration.
The reduction in enumeration is achieved via 
a {\em pruning} strategy which reduces the search space tremendously.
The algorithm works by
performing at each step $i$, a {\em depth first search} from vertex $v_i$, 
where the goal is to find the largest clique containing the vertex $v_i$.
At each {\em depth} of the search, the algorithm compares the number of remaining
vertices that could potentially constitute a clique containing vertex $v_i$
against the size of the largest clique encountered thus far.
If that number is found to be smaller, the algorithm backtracks (search is pruned).

\"{O}sterg{\aa}rd \cite{ostergard} devised an algorithm that incorporated an additional 
pruning strategy to the one by Carraghan and Pardalos.
The opportunity for the new pruning strategy is created by {\em reversing} the order in which the search is done by
the Carraghan-Pardalos algorithm. This allows for an additional pruning with the help of
some auxiliary book-keeping. 
Experimental results in \cite{ostergard} showed that the \"{O}sterg{\aa}rd
algorithm is faster than the Carraghan-Pardalos algorithm on random and 
DIMACS benchmark graphs \cite{dimacs}.
However, the new pruning strategy used in this algorithm is intimately tied to the order in which vertices are processed, introducing an inherent sequentiality into the algorithm.

A number of existing branch-and-bound algorithms for maximum clique
also use a vertex-coloring of the graph to obtain an upper bound on the maximum clique. 
A popular and recent method based on this idea is 
the MCQ algorithm of Tomita and Seki \cite{citeulike:7905505}. 
More recently, Konc and Jane\v{z}i\v{c} \cite{konc2007improved} presented 
an improved version of MCQ, known as MaxCliqueDyn (with the variants MCQD and MCQD+CS), that involves the use of tighter, computationally more expensive upper bounds 
applied on a fraction of the search space. Another improved version of MCQ is BBMC by San Segundo et al. \cite{SanSegundo}, which makes use of bit strings to sort vertices in constant time as well as to compute graph transitions and bounds efficiently.



%% file: algorithms.tex
\section{The New Algorithms}
\label{sec:algorithms}
We describe in this section new algorithms that overcome the shortcomings mentioned earlier; 
the new algorithms use additional pruning strategies, maintain simplicity, and 
avoid a sequential computational order.
We begin by first introducing the following notations.
We identify the $n$ vertices of the input graph $G=(V,E)$ as $\{v_1, v_2, \ldots, v_n\}$.  
The set of vertices adjacent to a vertex $v_i$, the set of its neighbors, is denoted by $N(v_i)$.
And the degree of the vertex $v_i$, the cardinality of $N(v_i)$, is 
denoted by $d(v_i)$. In our algorithm, the degree is computed once for each vertex at the beginning.

\subsection{The Exact Algorithm}
\label{subsec:exact}

Recall that the maximum clique in a graph can be found by computing the largest clique containing each vertex and picking the largest among these. 
A key element of our exact algorithm is that during the search for the largest clique containing a given vertex, vertices that cannot form cliques larger than the current maximum 
clique are {\em pruned}, in a hierarchical fashion. 
The method is outlined in detail in Algorithm \ref{alg:mClq}. 
Throughout, the variable $max$ stores the size of the maximum clique found 
thus far. Initially it is set to be equal to the lower bound $lb$ provided as an input parameter.
It gives the maximum clique size when the algorithm terminates.

\begin{figure}
\hrule
\vspace{6pt}
\begin{spacing}{\algospacing}
{
{\small
\captionof{algorithm}{{\protect\small Algorithm for finding the maximum clique of a given graph.
{\it Input}: Graph $G = \left (V, E\right )$, lower bound on clique $lb$ (default, 0). {\it Output}: Size of maximum clique.}}\label{alg:mClq}
\vspace{-5pt}
\hrule
\vspace{6pt}

\noindent\begin{minipage}{.5\textwidth}
\vspace{-12pt}
\begin{algorithmic}[1]
\Procedure {MaxClique}{$G=\left (V,E\right )$, $lb$}
\State $max \leftarrow lb$
\For{$i:1$ to $n$}
\If{$d(v_i) \ge max$} \Comment{{\footnotesize Pruning 1}} \label{pr1}
\State $U \leftarrow \emptyset$
\For{each $v_j \in N(v_i)$}
\If{$j > i$} \Comment{{\footnotesize Pruning 2}} \label{prOld}
\If{$d(v_j) \ge max$} \Comment{{\footnotesize Pruning 3}} \label{pr2}
\State $U \leftarrow U \cup \{v_j\}$ 
\EndIf
\EndIf
\EndFor

\State \textsc{Clique}$(G, U, 1)$ \label{subCall}
\EndIf

\EndFor
\EndProcedure
\end{algorithmic}
\end{minipage}%
\begin{minipage}{.5\textwidth}
\captionof*{algorithm}{{\em Subroutine}}
\vspace{-6pt}
\begin{algorithmic}[1]
\Procedure {Clique}{$G=\left (V,E\right )$, $U$, $size$}

\If{$U = \emptyset$}
\If{$size > max$} 
\State $max \leftarrow size$ \label{critical}
\EndIf
\State {\bf return}
\EndIf

\While{$\left|U\right| > 0$}
\If{$size + \left|{U}\right| \le max$} \Comment{{\footnotesize Pruning 4}} \label{pr3}
\State {\bf return} 
\EndIf

\State Select any vertex $u$ from $U$ 

\State $U \leftarrow U \setminus \{u\} $
\State $N'(u):= \{w | w \in N(u) \wedge d(w) \ge max\}$  \Comment{{\footnotesize Pruning 5}} \label{pr4}
\State \textsc{Clique}$( G, U \cap N'(u), size + 1)$
\EndWhile

\EndProcedure
\end{algorithmic}
\end{minipage}
}
}
\end{spacing}
\vspace{8pt}
\hrule
\end{figure}

To obtain the largest clique containing a vertex $v_i$, 
it is sufficient to consider only the neighbors of $v_i$. 
The main routine {\sc MaxClique} thus generates  for each 
vertex $v_i \in V$ a set $U \subseteq N(v_i)$ 
(neighbors of $v_i$ that survive pruning) and calls the subroutine \clq\ on $U$.  
The subroutine \clq\ goes through every relevant clique containing $v_i$ 
in a recursive fashion and returns the largest.
We use $size$ to maintain the size of the clique found at any point through the recursion.
Since we start with a clique of just one vertex, the value of $size$ is set to one initially, 
when \clq\ is called (Line \ref{subCall}, {\sc MaxClique}).

Our algorithm consists of several pruning steps.
Pruning 1 (Line \ref{pr1}, {\sc MaxClique}) 
filters vertices having strictly fewer neighbors than the size of the maximum clique already computed. These vertices can be ignored, since even if a clique were to be found, its size would not be larger than $max$.
While forming the neighbor list $U$ for a vertex $v_i$, we include only those of $v_i$'s 
neighbors for which the largest clique containing them has not been found 
(Pruning 2; Line \ref{prOld}, {\sc MaxClique}), 
to avoid recomputing previously found cliques.  
Pruning 3 (Line \ref{pr2}, {\sc MaxClique})
excludes vertices $v_j \in N(v_i)$  that have degree less than the current value of $max$, since any such vertex could not form a clique of size larger than $max$.
Pruning 4 (Line \ref{pr3}, {\sc Clique})
checks for the case where even if all vertices of $U$ were added to get a clique, its size would not exceed that of the largest clique encountered so far in the search, $max$. 
Pruning 5 (Line 11, {\sc Clique})
reduces the number of comparisons needed to generate the intersection set in Line 12.
Note that the routine \clq\ is similar to the 
Carraghan-Pardalos algorithm \cite{pardalos}; Pruning 5 accounts for the main difference.
Also, Pruning 4 is used in most existing algorithms, whereas Prunings 1, 2, 3 and 5 are not.

\subsection{The Heuristic}
\label{subsec:heuristic}

\begin{figure}
\vspace{10pt}
\hrule
\vspace{6pt}
\begin{spacing}{\algospacing}
{
{\small
\captionof{algorithm}{{\protect\small Heuristic for finding the maximum clique in a graph.
{\it Input}: Graph $G = \left (V, E\right )$. {\it Output}: Approximate size of maximum clique.}}\label{alg:mClqHeu}
\vspace{-5pt}
\hrule
\vspace{6pt}

\noindent\begin{minipage}{.5\textwidth}
\vspace{12pt}
\begin{algorithmic}[1]
\Procedure {MaxCliqueHeu}{$G=\left (V,E\right )$}
\For{$i:1$ to $n$}
\If{$d(v_i) \ge max$}
\State $U \leftarrow \emptyset$
\For{each $v_j \in N(v_i)$}
\If{$d(v_j) \ge max$} 
\State $U \leftarrow U \cup \{v_j\}$ 
\EndIf
\EndFor

\State \textsc{CliqueHeu}$(G, U, 1)$
\EndIf

\EndFor
\EndProcedure
\end{algorithmic}
\end{minipage}%
\begin{minipage}{.5\textwidth}
\captionof*{algorithm}{{\em Subroutine}}
\vspace{-7pt}
\label{alg:clqHeu}
\begin{algorithmic}[1]
\Procedure {CliqueHeu}{$G=\left (V,E\right )$, $U$, $size$}

\If{$U = \emptyset$}
\If{$size > max$}
\State $max \leftarrow size$
\EndIf
\State {\bf return}
\EndIf


\State Select a vertex $u \in U$ of maximum degree in $G$ \label{maxDsel}
\State $U \leftarrow U \setminus \{u\} $
\State $N'(u):= \{w | w \in N(u) \wedge d(w) \ge max\}$  \label{pr4}
\State \textsc{CliqueHeu}$( G, U \cap N'(u), size + 1)$


\EndProcedure

\end{algorithmic}
\end{minipage}
}
}
\end{spacing}
\vspace{8pt}
\hrule
\end{figure}

The exact algorithm examines all relevant cliques containing every vertex.
Our heuristic, shown in Algorithm \ref{alg:mClqHeu}, considers only {\em one} neighbor
with {\em maximum degree} at each step instead of recursively considering {\em all} neighbors 
from the set $U$, and thus is much faster. The vertex with maximum degree is chosen
for this intuitive reason: in a relatively fairly connected subgraph, a vertex with the maximum degree
is more likely to be a member of the largest clique containing that vertex
compared to any other.

\subsection{Complexity}
\label{subsec:complexity}

The exact algorithm, Algorithm \ref{alg:mClq}, examines for every vertex $v_i$ all candidate cliques containing the vertex $v_i$ in its search for the largest clique. Its time complexity is exponential in the worst case. The heuristic, Algorithm \ref{alg:mClqHeu}, loops over the $n$ vertices, each time possibly
calling the subroutine \clqh, which effectively is a loop that runs until the set $U$ is empty. 
Clearly, $|U|$ is bounded by the max degree $\Delta$ in the graph.  
The subroutine also includes the computation of a neighbor list, whose runtime is bounded by 
$O(\Delta)$.
Thus, the time complexity of the heuristic is bounded by $O(n\cdot \Delta^{2})$.

\subsection{Graph Data Structure}
\label{subsec:implementation}

\begin{figure}
  \centering
    \includegraphics[scale=0.28]{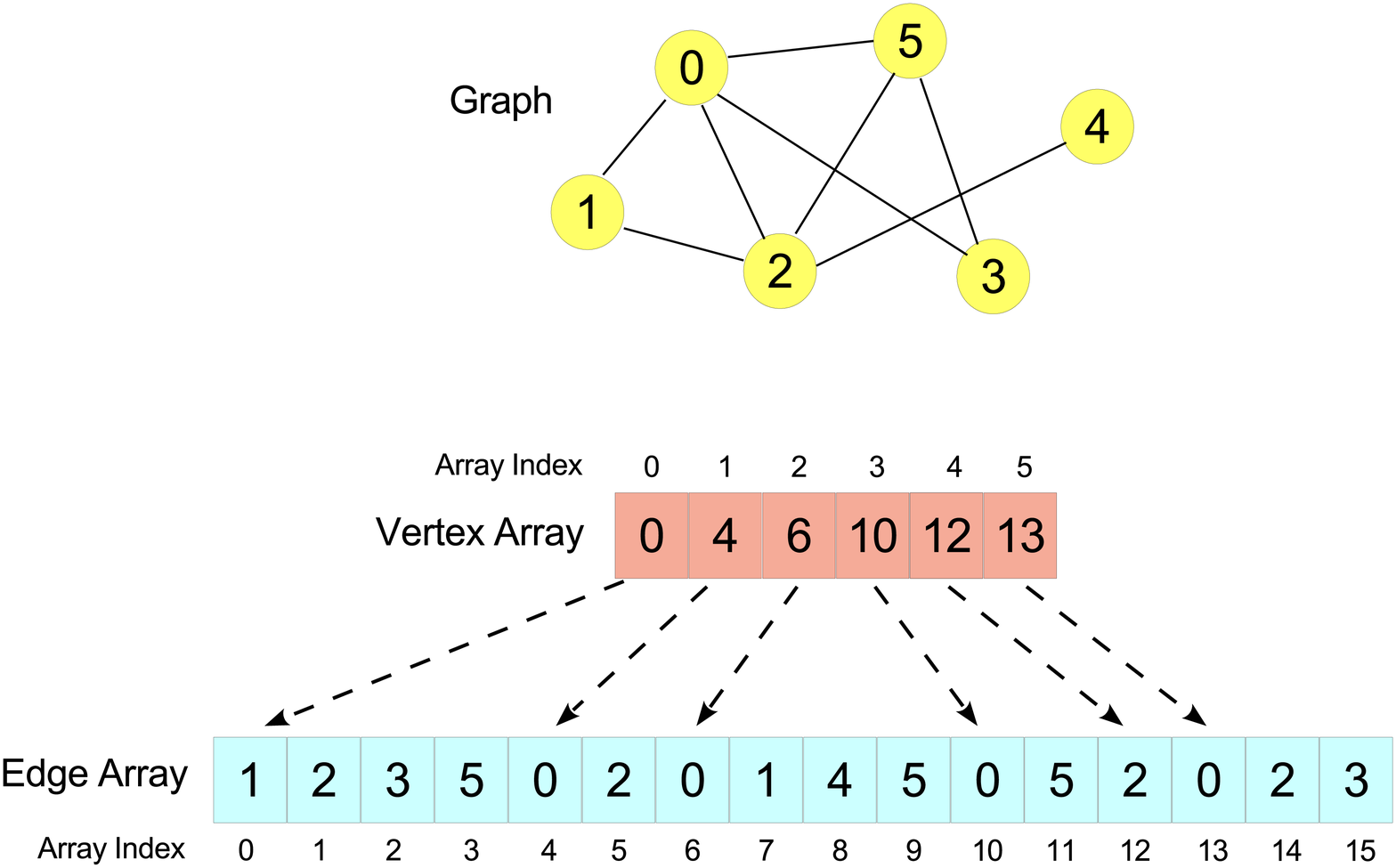}
\caption{The adjacency list data structure used in our implementation shown for a sample graph. Each element in the vertex array stores the index to the starting element of its neighbor list in the edge array.}
\label{fig-implementation}
\end{figure}

Our implementation uses a simple adjacency list representation to store the graph. This is done by maintaining two arrays. Given a graph $G=(V,E)$, with its vertices numbered from $0$ to $|V|-1$, the edge array maintains the concatenated list of sorted neighbors of each vertex. The size of this array is $2|E|$. The vertex array is $|V|$ elements long, one for each vertex in sequential order, and each element points (stores the array index) to starting point of its neighbor list in the edge array. 
Figure \ref{fig-implementation} shows our representation and data structure used for a sample graph.

%% file: experiments.tex
\section{Experimental Evaluation}
\label{sec:experiments}

We present in this section results comparing the performance of our algorithm
with other existing algorithms. Our experiments were performed on a Linux workstation running 64-bit Red Hat Enterprise Linux Server release 6.2 with a 2GHz Intel Xeon E7540 processor with 32GB of main memory. Our implementation is in C++, compiled using gcc version 4.4.6 with -O3 optimization.

\subsection{Test Graphs}

\begin{table}[!h]
\centering
\caption{Overview of real-world graphs in the testbed (UF, Pajek and Stanford datasets) and their origins.}
\label{tab:real-graphs}
\begin{tabular}{ll}
\hline \hline
{\bf Graph} & {\bf Description} \\ \hline \hline
{\it cond-mat-2003} \cite{Newman06042004} & A collaboration network of scientists posting preprints 
on \\ & the condensed matter archive at www.arxiv.org in the period \\ 
& between January 1, 1995 and June 30, 2003. \\
{\it email-Enron} \cite{Leskovec:2005:GOT:1081870.1081893} & A communication network representing
email exchanges. \\
{\it dictionary28} \cite{pajek2006} & Pajek network of words. \\ 
{\it Fault\_639} \cite{Ferronato20083922} & A structural problem discretizing a faulted gas reservoir with \\
& tetrahedral Finite Elements and triangular Interface Elements. \\
{\it audikw\_1} \cite{Davis97theuniversity} & An automotive crankshaft model of TETRA elements. \\ 
{\it bone010} \cite{vanRietbergen199569} &
A detailed micro-finite element model of bones representing \\
& the porous bone micro-architecture. \\
{\it af\_shell} \cite{Davis97theuniversity}  & A sheet metal forming simulation network. \\ 
{\it as-Skitter} \cite{Leskovec:2005:GOT:1081870.1081893} & An Internet topology graph from trace routes run daily in 2005. \\  
{\it roadNet-CA} \cite{Leskovec:2005:GOT:1081870.1081893} & A road network of California.
Nodes represent intersections \\ & and endpoints and edges represent the roads connecting them. \\  
{\it kkt\_power} \cite{Davis97theuniversity} & An Optimal Power Flow (nonlinear optimization) network. \\\hline
{\it foldoc} \cite{foldoc} & A searchable dictionary of terms related to computing. \\
{\it eatRS} \cite{eatRS} & The Edinburgh Associative Thesaurus, a set of word association \\ &
norms showing the counts of word association as collected \\ & from subjects. \\
{\it hep-th} \cite{hep-th-kdd}	& Citation data from KDD Cup 2003, a knowledge discovery and\\ & data mining
competition held in conjunction with the Ninth\\ & ACM SIGKDD Conference. \\
{\it patents} \cite{hall2001}	& Data set containing information on almost 3 million U.S.\\ & patents granted
between January 1963 and December 1999, and\\ & all citations made to
these patents between 1975 and 1999. 	\\
{\it days-all} \cite{corman2002}	& Reuters terror news network obtained from the CRA networks\\ & produced
by Steve Corman and Kevin Dooley at Arizona\\ & State University.	\\\hline
{\it roadNet-TX} \cite{Leskovec:2005:GOT:1081870.1081893}	& Road network of Texas.	\\
{\it amazon0601} \cite{leskovec2007}	& Amazon product co-purchasing network from June 1 2003. \\
{\it email-EuAll} \cite{leskovec2007-2}	& Email network from a EU research institution.	\\
{\it web-Google} \cite{web-google}	& Web graph released by Google in 2002 as a part of
Google\\ & Programming Contest.	\\
{\it soc-wiki-Vote} \cite{leskovec2010}	& Wikipedia who-votes-on-whom network.	\\
{\it soc-slashdot0902} \cite{Leskovec:2005:GOT:1081870.1081893}	& Slashdot social network from November 2008.\\
{\it cit-Patents} \cite{hall2001}	& Citation network among US Patents.	\\
{\it soc-Epinions1} \cite{richardson2003}	&	Who-trusts-whom network of Epinions.com. \\
{\it soc-wiki-Talk} \cite{leskovec2010}	&	Wikipedia talk (communication) network. \\
{\it web-berkstan} \cite{Leskovec:2005:GOT:1081870.1081893}	& Web graph of Berkeley and Stanford.	\\\hline\hline

\end{tabular}
\end{table}

Our testbed consists of 91 graphs, grouped in three categories:\\
\begin{enumerate*}[label=\textbf{\arabic*})]
\item {\bf Real-world graphs. } 
Under this category, we consider 10 graphs downloaded from the 
University of Florida (UF) Sparse Matrix Collection  \cite{Davis97theuniversity}, 5 graphs from Pajek data sets \cite{pajek2006}, and 10 graphs from the Stanford Large Network Dataset Collection \cite{stanford_dataset}.
The graphs originate
from various real-world applications. 
Table~\ref{tab:real-graphs} gives a quick overview of the graphs and their origins.\\

\item {\bf Synthetic Graphs. } 
In this category we consider 15 graphs generated using 
the R-MAT algorithm \cite{Chakrabarti:2006:GML:1132952.1132954}. The graphs
are subdivided in three categories depending on the structures they represent:\\

\begin{enumerate*}[label=\textbf{\alph*})]
\item {\bf Random graphs} (5 graphs) -- Erd\H{o}s-R\'{e}nyi random  graphs generated using R-MAT with the parameters (0.25, 0.25, 0.25, 0.25).  We denoted these with prefix {\it rmat\_er}.\\
\item {\bf Skewed Degree, Type 1 graphs} (5 graphs) -- graphs generated using R-MAT with the parameters (0.45, 0.15, 0.15, 0.25). These are denoted with prefix {\it rmat\_sd1}.\\
\item {\bf Skewed Degree, Type 2 graphs} (5 graphs) --  graphs generated using R-MAT with the parameters (0.55, 0.15, 0.15, 0.15). These are denoted with prefix {\it rmat\_sd2}.\\
\end{enumerate*}

\item {\bf DIMACS graphs. } 
This last category consists of 51 graphs selected from the Second DIMACS Implementation Challenge \cite{dimacs}. Among these, 5 graphs are considered for discussion in the next few sections, while the results for the rest are reported in Table \ref{tab:dimacs} in the Appendix.
\end{enumerate*}

The DIMACS graphs  are an established benchmark for the maximum
clique problem, but they are of rather limited size and variation. 
In contrast, the real-work networks included  in category 1 of the testset
and the synthetic (RMAT) graphs in category 2
represent a wide spectrum of large graphs posing varying degrees of difficulty for the algorithms. 
The {\it rmat\_er} graphs have {\it normal} degree distribution, whereas the {\it rmat\_sd1} and {\it rmat\_sd2} graphs have skewed degree distributions and contain many dense local subgraphs.
 The {\it rmat\_sd1} and {\it rmat\_sd2} graphs differ primarily in the magnitude of maximum vertex degree they contain; the {\it rmat\_sd2} graphs have much higher maximum degree. 
Table \ref{tab:struc-graphs} lists basic structural information (the number of vertices, 
number of edges and the maximum degree) about 45 of the test graphs (25 real-world, 15 synthetic and 5 DIMACS).

\input{graphs_properties}

\subsection{Algorithms for Comparison}
The algorithms we consider for comparison are the ones by: 
\begin{itemize}
\itemsep0em
\item Carraghan-Pardalos \cite{pardalos}. We used our own implementation of this algorithm.
\item \"{O}sterg{\aa}rd algorithm \cite{ostergard}. We used the publicly available {\it cliquer} source code \cite{cliquer}.
\item Konc and Jane\v{z}i\v{c} \cite{konc2007improved}. We used the code {\it MaxCliqueDyn} or MCQD, available at \\{\small \url{http://www.sicmm.org/~konc/maxclique/}}. Among the variants available in MCQD, we report results on MCQD+CS (which uses improved coloring and dynamic sorting), since it is the best-performing variant. The {\it MaxCliqueDyn} code was not capable of handling large input graphs and had to be aborted for many instances. For those that it successfully ran, we had to first modify the graph reader to make it able to handle graphs with multiple connected components.
\item Tomita and Seki (MCQ) \cite{citeulike:7905505}.
\item Tomita et al. (MCS) \cite{walcom}.
\item San Segundo et al. (BBMC) \cite{SanSegundo}.
\end{itemize}
For MCQ, MCS, and BBMC, we used the publicly available Java implementation, MCQ1, MCSa1, and BBMC1 respectively, by Prosser \cite{prosser2012}  available at {\small \url{http://www.dcs.gla.ac.uk/~pat/maxClique/}}. These implementations failed to run due to memory limitations in spite of making available 20 GB of memory for almost all (except two) of the larger data sets. Hence, timings are reported only for DIMACS graphs. 

In addition to comparing with the above mentioned algorithms, for the Pajek and Stanford data sets, we also provide comparison of the timing results of our algorithm with the maximal clique enumeration algorithm by Epstein \& Strash \cite{sea}. For this, we directly quote numbers published in \cite{sea}, and the results are listed in Table~\ref{tab:pajek_stanford-es} in the Appendix.
It should to be kept in mind that maximal clique enumeration in general is a harder problem than maximum clique finding, and that these experiments have been performed on different test environment. Therefore the runtime results should be understood in a sense qualitatively.

\subsection{Results}
\label{sec:exp-results}

\input{results_timings}
\input{new_timings}

Table~\ref{tab:timings} shows the size of the maximum clique ($\omega$) and the runtimes  of our exact algorithm (Algorithm 1) and the algorithms of Caraghan and Pardalos (CP), 
\"{O}sterg{\aa}rd ({\it cliquer}) 
and Konc and Jane\v{z}i\v{c}  (MCQD+CS) for all the graphs in the testbed except for 
the Pajek and Stanford test graphs;
results on the Pajek and Stanford graphs are presented in Table~\ref{tab:pajek_stanford}.
The last two columns in Table~\ref{tab:timings} show the results of our 
heuristic (Algorithm 2)---the size of the maximum clique 
returned ($\omega_{A_2}$)  and its runtime ($\tau_{A_2}$). 
The columns labeled $P1$, $P2$, $P3$ and $P5$ list the number of 
vertices/branches pruned in the respective pruning steps of Algorithm 1.
Data on Pruning 4 is omitted since that pruning is used by all of the algorithms compared in the table. The displayed pruning numbers have been rounded  (K stands for $10^3$, M for $10^6$ and B for $10^9$);
the exact numbers can be found in Table \ref{tab:prunings} in the Appendix.

In Table \ref{tab:timings}, the fastest runtime for each instance is indicated with boldface. 
An asterisk (*) indicates that an algorithm did not terminate within 25,000 seconds for a particular
instance. A hyphen (-) indicates that the publicly available implementation 
(the {\it MaxCliqueDyn} code) had to be aborted because the input graph was too large 
for the implementation to handle.
For the graph {\it rmat\_sd2\_5}, none of the algorithms computed the maximum clique size in 
a reasonable time; the entry there is marked with N, standing for  ``Not Known''.

We discuss in what follows our observations from this table
for the exact algorithm and the heuristic.

\subsubsection{Exact algorithms}
\label{sec:exp-exact}

As expected, our exact algorithm gave the same size of maximum clique as the other
three algorithms for all test cases. 
In terms of runtime,  its relative performance compared to the other three varied
in accordance with the advantages afforded by the various pruning steps.  

Vertices that are discarded by Pruning 1 are skipped in the main loop of the algorithm, and the largest cliques containing them are not computed. Pruning 2 avoids re-computing previously computed cliques in the neighborhood of a vertex. In the absence of Pruning 1, the number of vertices pruned by Pruning 2 would be bounded by the number of edges in the graph (note that this is more than the total number of vertices in the graph). While Pruning 3 reduces the size of the input set on which the maximum clique is to be computed, Pruning 5 brings down the time taken to generate the intersection set in Line 12 of the subroutine. 
Pruning 4 corresponds to back tracking. Unlike Pruning steps 1, 2, 3 and 5, Pruning 4
is used  by all three of the other algorithms in our comparison. The primary strength of our algorithm is its ability to take advantage of pruning in multiple steps in a hierarchical fashion, allowing for opportunities for one or more of the steps to be triggered and positively impact the performance.

\begin{figure}
  \centering
    \includegraphics[scale=0.5]{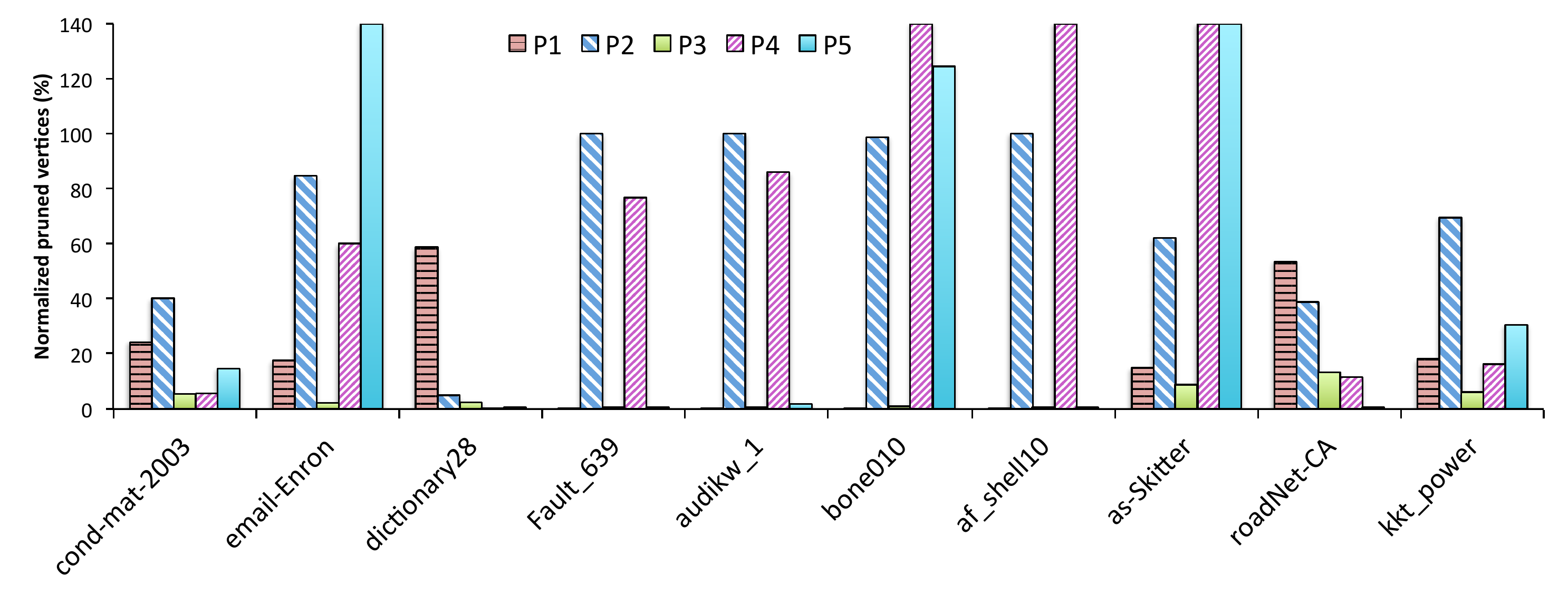}
\caption{Number of ``pruned" vertices in the various pruning steps normalized
by the number of edges in the graph (in percents) for the UF collection graphs (we cut few bars reachining 140\% as their correspnding values are much higher).}
\label{fig-pruningplot}
\end{figure}

In Figure~\ref{fig-pruningplot} we show the number of vertices discarded by all
the  pruning steps of the exact algorithm normalized by the total number of edges
in a graph for the real-world graphs in Table \ref{tab:timings}. We cut few bars reachining
140\% as their correspnding values are much higher.
In the Appendix, we provide a complete tabulation of the raw numbers for the pruned vertices in all the steps for all the graphs in the testbed. It can be seen for these graphs pruning steps 2 and 5 in particular discard a large percentage of vertices, potentially resulting in large runtime savings. The general behavior of the pruning steps Pruning 1, 2, 3 and 5 for the synthetic graphs {\em rmat\_er} and {\em rmat\_sd1} was observed to be somewhat similar to that depicted in Figure~\ref{fig-pruningplot} for the real-world graphs. In contrast, for the DIMACS graphs, the number of vertices pruned in steps Pruning 1, 3 and 5 were observed to be zero; the numbers in the step Pruning 2 were non-zero but relatively modest.

As a result  of the differences seen in the effects of the pruning steps, as discussed below,
the runtime performance of our algorithm (seen in Table \ref{tab:timings}) compared
to the other three algorithms varied in accordance with the difference in the structures represented 
by the different categories of graphs in the testbed.

{\bf Real-world Graphs. }
For most of the graphs in this category (Table \ref{tab:timings}), it can be seen that our algorithm runs several orders of magnitude faster than the other three, mainly due to the large amount of pruning the algorithm attained. These numbers also illustrate the great benefit of hierarchical pruning. 
For the graphs {\em Fault\_639}, {\em audikw\_1} and {\em af\_shell10}, 
Prunings 1, 3 and 5 have only minimal impact,
whereas Pruning 2 makes a big difference resulting in impressive runtimes. 
The number of vertices pruned in steps Pruning 1 and 3 varied among the 
graph {\em within} the
category, ranging from 0.001\% for {\it af\_shell} to a staggering 97\% for {\it as-Skitter} 
for the step Pruning 1. For the graphs in Table \ref{tab:pajek_stanford} as well, one can observe that our algorithm performs much better than the others.

{\bf Synthetic Graphs. }
For the synthetic graph types {\it rmat\_er} and {\it rmat\_sd1}, our algorithm clearly outperforms 
the other three by a few orders of magnitude in all cases. 
This is also primary due to the high number of vertices discarded by the new pruning steps. 
In particular, for {\it rmat\_sd1} graphs, between 30 to 37\% of the vertices are pruned just in the step Pruning 1. 
For the {\it rmat\_sd2} graphs, which have relatively larger maximum clique and higher maximum degree than the {\it rmat\_sd1} graphs, our algorithm is observed to be faster than 
CP but slower than {\em cliquer}. 

{\bf DIMACS Graphs. }
The runtime of our exact algorithm for the DIMACS graphs is 
in most cases comparable to that of CP and higher than that of {\it cliquer}
and MCQD+CS. 
For these graphs, only Pruning 2 was found to be effective, 
and thus the performance results agree with one's expectation. 
The timings on a much larger collection of DIMACS graphs are presented in Table \ref{tab:dimacs} in the Appendix.

It is to be noted that the DIMACS graphs are intended to serve as challenging test cases for the maximum clique problem, and graphs with such high edge densities and low vertex count are rare in practice. 
Most of these have between 20 to 1024 vertices with an average edge density of roughly 0.6, 
whereas, most real world graphs are often very large and sparse
as exemplified by the real-world graphs in our test bed.
Other good examples of instances with similar nature include Internet topology graphs \cite{Faloutsos:1999:PRI:316188.316229}, the web graph \cite{kumar:extracting}, and social network graphs \cite{Domingos:2001:MNV:502512.502525}.

\subsubsection{The Heuristic}
\label{sec:exp-heuristic}

It can be seen that our heuristic runs several orders of magnitude faster than our exact algorithm,
while delivering either optimal or very close to optimal solution.
It gave the optimal solution on 25 out of the 30 test cases.
On the remaining 5 cases where it was suboptimal, its accuracy ranges from 83\% to 99\% (on average 93\%).

Additionally, we run the heuristic by choosing a vertex randomly in Line \ref{maxDsel} of Algorithm \ref{alg:mClqHeu} instead of the one with the maximum degree. We observe that on average, the solution is optimal only for less than $40\%$ of the test cases compared to 83\% when selecting the maximum degree vertex.

\begin{SCfigure}
  \centering
    \includegraphics[width=7.8cm]{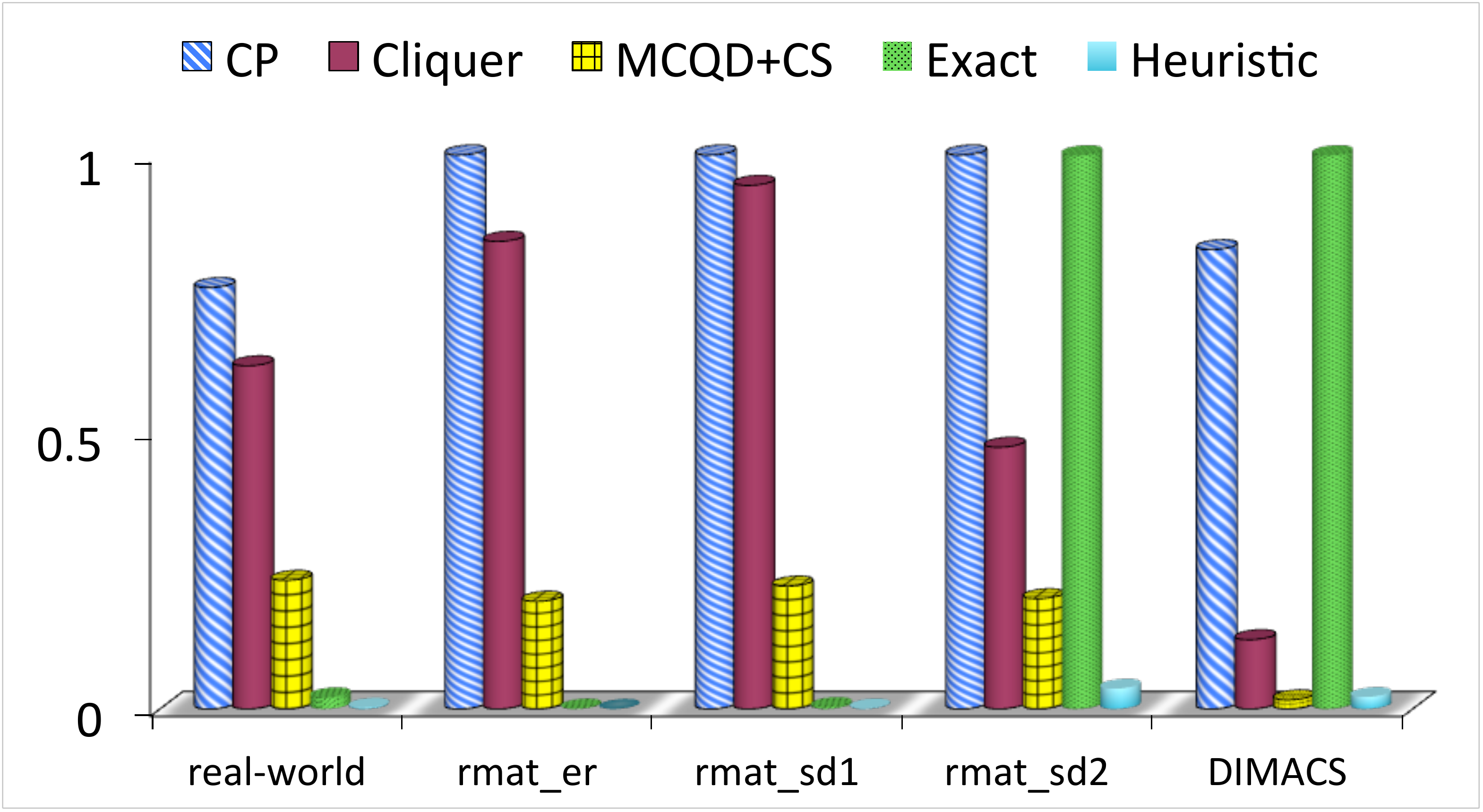}
  \caption{Runtime (normalized, mean) comparison between various algorithms. For each category of graph, first, all runtimes for each graph were normalized by the runtime of the slowest algorithm for that graph, and then the mean was calculated for each algorithm. Graphs were considered only if the runtimes for at least three algorithms was less than the 25,000 seconds limit set.}
\label{fig-timeplot}
\end{SCfigure}

\subsubsection{Further Analysis}
\label{sec:exp-further}

Figure~\ref{fig-timeplot} provides an aggregated visual summary of the runtime trends of
the various algorithms across the five categories of graphs in the testbed. 

\begin{figure}
  \centering
    \includegraphics[scale=0.67]{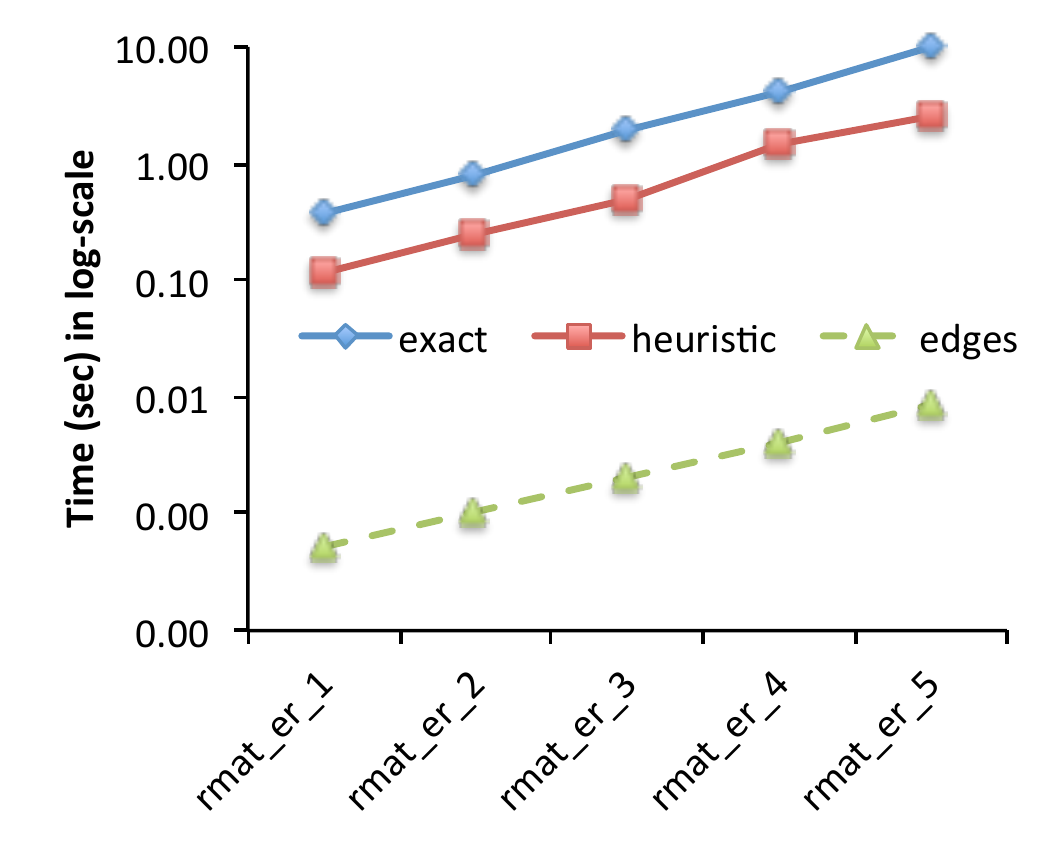}
    \includegraphics[scale=0.67]{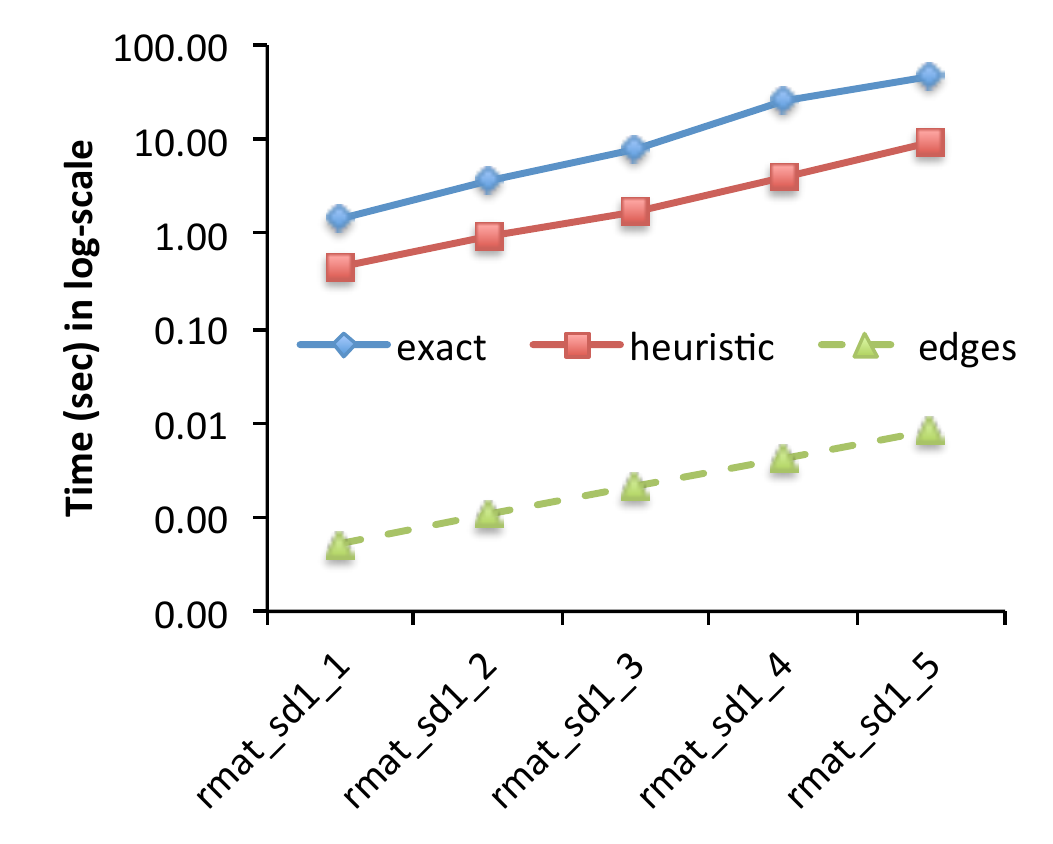}\\
    
    \includegraphics[scale=0.67]{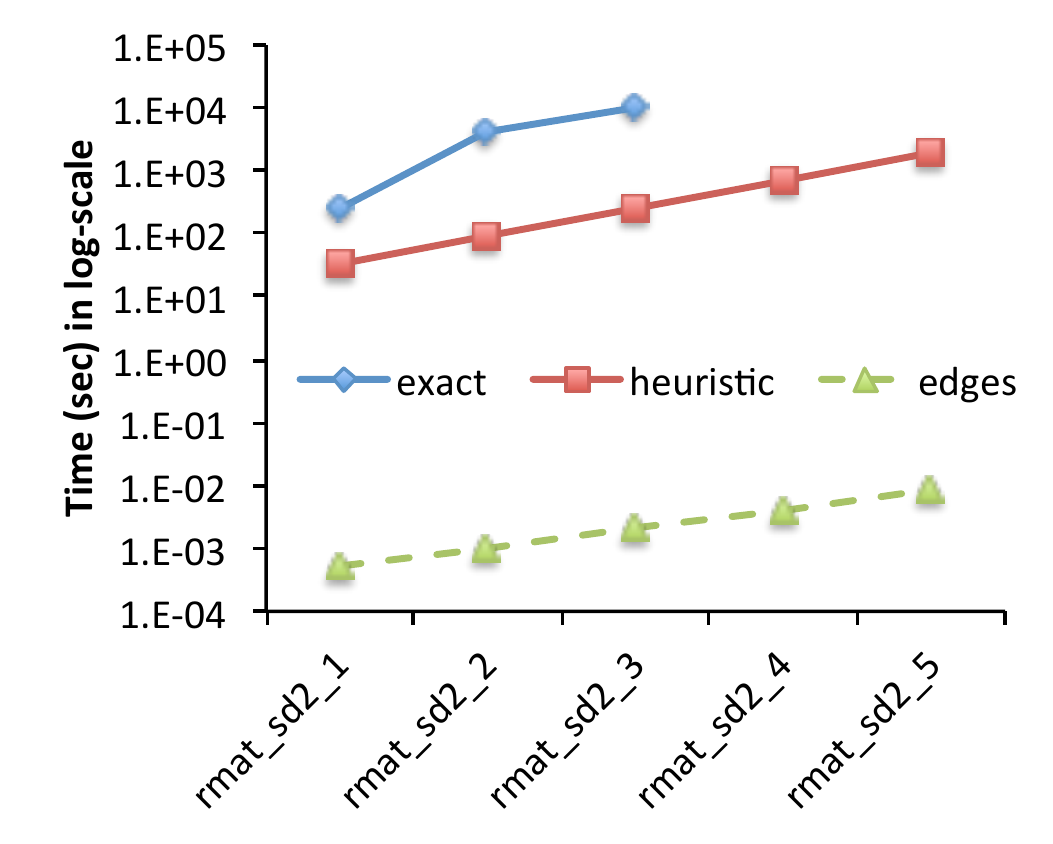}
    \includegraphics[scale=0.67]{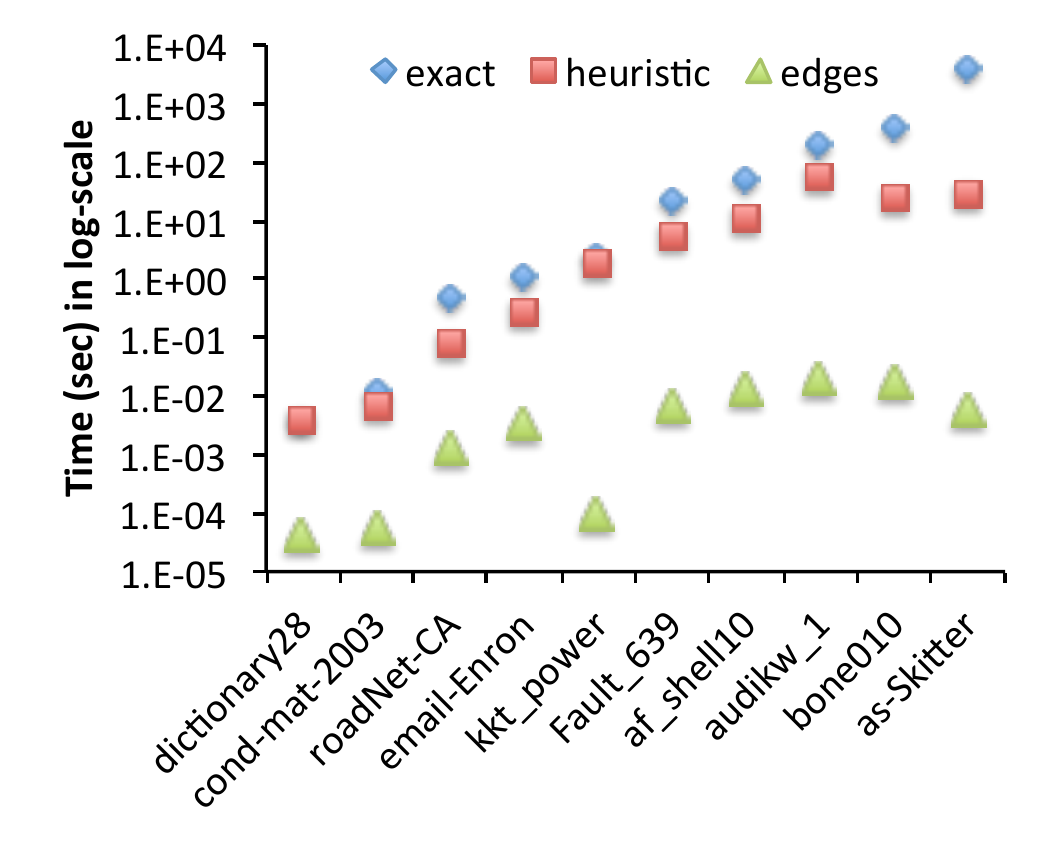}
    
 \caption{Run time plots of the new exact and heuristic algorithms. The third curve, labeled
 {\em edges}, shows the quantity number of edges in the graph divided by the clock 
 frequency of the computing platform used in the experiment. 
 }
\label{fig-runtimeplots}
\end{figure}

To give a sense of runtime growth rates, we provide in Figure~\ref{fig-runtimeplots} plots of the 
runtime of the new exact algorithm and the heuristic for the synthetic and real-world graphs 
in the testbed. In addition to the curves corresponding to the runtimes of the
{\em exact} algorithm and the {\em heuristic}, the figures also include a curve corresponding to
the number of {\em edges} in the graph divided by the clock frequency of the computing
platform used in the experiment. This curve is added to facilitate comparison between
the growth rate of the runtime of the algorithms with that of a linear-time (in the size of the graph) growth rate. 
It can be seen that the runtime of the heuristic by and large grows 
somewhat linearly with the size of a graph. The exact algorithm's runtime, which is orders of
magnitude larger than the heuristic, exhibited a similar growth behavior for these test-cases
even though its worst-case complexity suggests exponential growth in the general case.

%% file: graphs_properties.tex
\begin{table}[t]
\centering
\caption{Structural properties---the number of vertices $|V|$; the umber of edges $|E|$; and the maximum degree $\Delta$---of the graphs $G$ in the testbed.
The graphs on the left side are graphs from the UF collection, Pajek data sets, and Stanford Large Dataset collection. On the right are the   
RMAT graphs; and the DIMACS Challenge graphs.}  
\label{tab:struc-graphs}
\begin{tabular}{l@{\hspace{5pt}}r@{\hspace{5pt}}r@{\hspace{5pt}}r@{\hspace{5pt}}|@{\hspace{5pt}}l@{\hspace{5pt}}r@{\hspace{5pt}}r@{\hspace{5pt}}r}

\toprule\toprule

$G$ & $|V|$ & $|E|$ & $\Delta$ & $G$ & $|V|$ & $|E|$ & $\Delta$ \\ \hline \hline
{\it cond-mat-2003} & 31,163	& 120,029	 & 202 &	{\it rmat\_sd1\_1} &    131,072 &    1,046,384 & 407           \\ \vspace*{\rowspace}
{\it email-Enron} & 36,692	 & 183,831 &	1,383  &	{\it rmat\_sd1\_2} &    262,144 &    2,093,552 &   558    \\ \vspace*{\rowspace}
{\it dictionary28} & 	52,652 &	89,038 &	38  &		{\it rmat\_sd1\_3} &    524,288 &    4,190,376 &    618  \\ \vspace*{\rowspace}
{\it Fault\_639} &    638,802 &    13,987,881 &    317 &  {\it rmat\_sd1\_4} &    1,048,576 &    8,382,821 &  802    \\ \vspace*{\rowspace}
{\it audikw\_1} &    943,695 &    38,354,076 &    344  &	 {\it rmat\_sd1\_5} &    2,097,152 &    16,767,728 &    1,069   \\ \vspace*{\rowspace}
{\it bone010} &    986,703 &    35,339,811 &    80 & 
{\it rmat\_sd2\_1} &    131,072 &    1,032,634 &    2,980        \\ \vspace*{\rowspace}
{\it af\_shell10} &    1,508,065 &    25,582,130 &    34 &  
{\it rmat\_sd2\_2} &    262,144 &    2,067,860 &    4,493      \\ \vspace*{\rowspace}
{\it as-Skitter} &    1,696,415 &    11,095,298 &  35,455 &
{\it rmat\_sd2\_3} &    524,288 &    4,153,043 &    6,342     \\ \vspace*{\rowspace}
{\it roadNet-CA} &    1,971,281 &    2,766,607 &    12 & 
{\it rmat\_sd2\_4} &    1,048,576 &    8,318,004 &    9,453      \\ \vspace*{\rowspace}	
{\it kkt\_power} &    2,063,494 &    6,482,320 &    95  &   
{\it rmat\_sd2\_5} &    2,097,152 &    16,645,183 &    14,066    \\

\midrule
\vspace*{\rowspace}

{\it foldoc}				&	13,356	&	91,470		&	728		&
{\it rmat\_er\_1} &    131,072 &    1,048,515 &    82 	 \\ \vspace*{\rowspace}  
{\it eatRS}				&	23,219	&	304,938		&	1090		&
{\it rmat\_er\_2} &    262,144 &    2,097,104 &    98 	 \\ \vspace*{\rowspace}
{\it hep-th}				&	27,240	&	341,923		&	2411		&
{\it rmat\_er\_3} &    524,288 &    4,194,254 &    94  	 \\ \vspace*{\rowspace}
{\it patents}			&	240,547	&	560,943		&	212		&
{\it rmat\_er\_4} &    1,048,576 &    8,388,540 &    97  	 \\ \vspace*{\rowspace}
{\it days-all}			&	13,308	&	148,035		&	2265		&
{\it rmat\_er\_5} &    2,097,152 &    16,777,139 &    102 	 \\
\midrule
\vspace*{\rowspace}

{\it roadNet-PA}			&	1,088,092	&	1,541,898		&	9		&
{\it hamming6-4} & 64 &    704 &    22 	 \\ \vspace*{\rowspace}
{\it roadNet-TX}			&	1,393,383	&	1,921,660		&	12		&
{\it johnson8-4-4} & 70 &    1,855 &    53   \\ \vspace*{\rowspace} 
{\it amazon0601}		&	403,394	&	2,247,318		&	2752		&
{\it keller4} &    171 &    9,435 &    124  \\ \vspace*{\rowspace}   
{\it email-EuAll}			&	265,214	&	364,481		&	7636		&
{\it c-fat200-5} &    200 &    8,473 &    86  \\ \vspace*{\rowspace} 
{\it web-Google}		&	916,428	&	4,322,051		&	6332		&
{\it brock200\_2} &    200 &    9,876 &    114 \\ \vspace*{\rowspace}
{\it soc-wiki-Vote}		&	8,297	&	100,762		&	1065		&
 &     &     &     \\ \vspace*{\rowspace}
{\it soc-slashdot0902}	&	82,168	&	504,230		&	2252		&
 &     &     &     \\ \vspace*{\rowspace}
{\it cit-Patents}			&	3,774,768	&	16,518,947	&	793		&
 &     &     &     \\ \vspace*{\rowspace}
{\it soc-Epinions1}		&	75,888	&	405,740		&	3044		&
 &     &     &     \\ \vspace*{\rowspace}
{\it soc-wiki-Talk}		&	2,394,385	&	4,659,565		&	100029	&
 &     &     &     \\ \vspace*{\rowspace}
{\it web-berkstan}		&	685,230	&	6,649,470		&	84230	&
 &     &     &     \\ 
\bottomrule\bottomrule
\end{tabular}
\end{table}

%% file: results_timings.tex
\begin{table}[!hbt]

\small
\centering
\caption{Comparison of runtimes (in seconds) of algorithms {\it CP}~\cite{pardalos}, 
 {\it cliquer}~\cite{ostergard},  MCQD+CS~\cite{konc2007improved} and
and our new exact algorithm (${A_1}$) for the graphs in the testbed. An asterisk (*) indicates that the algorithm did not terminate within 25,000 seconds for a particular
instance. A hyphen (-) indicates that the publicly available implementation we used could not handle this instance due to its large size. Columns $P1$, $P2$, $P3$ and $P5$ list the number of vertices/branches pruned in steps Pruning 1, 2, 3 and 5 of our exact algorithm (K stands for the quantity $10^3$, M for $10^6$ and B for $10^9$). 
The column $\omega$ (second column) lists the maximum clique size in each graph, 
the column $\omega_{A_2}$ lists the clique size returned by our heuristic and the column
$\tau_{A_2}$ lists the heuristic's runtime.}
\label{tab:timings}
\begin{tabular}{l@{\hspace{6pt}}r@{\hspace{6pt}}|@{\hspace{6pt}}r@{\hspace{6pt}}r@{\hspace{6pt}}r@{\hspace{6pt}}r@{\hspace{6pt}}|@{\hspace{4pt}}r@{\hspace{4pt}}r@{\hspace{4pt}}r@{\hspace{4pt}}r@{\hspace{4pt}}|@{\hspace{6pt}}r@{\hspace{6pt}}r}

\toprule\toprule
           &  & & &  $\tau_{MCQD}$ & &	&	&	&	&& \\
Graph           	& $\omega$ & $\tau_{CP}$    & $\tau_{cliquer}$  & $_{+CS}$  &  $\tau_{A_1}$		& 	$P1$ 		&	$P2$		& 	$P3$ 		&	$P5$		 & $\omega_{A_2}$ &  $\tau_{A_2}$ \\
\hline \hline
{\it cond-mat-2003} 	& 	25 	& 	4.87 			&  	11.17		&	2.41		&	{\bf 0.011}		& 	29K 			&	48K			&	6,527 		& 	17K			&	25 		& 	$<$0.01 	\\
{\it email-Enron} 	& 	20 	& 	7.01			& 	15.08 		&	3.70		& 	{\bf 0.998}		& 	32K 			&	155K		&	4,060 		& 	8M			&	18 		& 	0.26	\\ 
{\it dictionary28} 	& 	26 	& 	7.70	 		&	32.74 		&	7.69		&	{\bf $<$0.01}	& 	52K			& 	4,353		&	2,114		& 	107			&	26 		&	$<$0.01	\\	
{\it Fault\_639}		&	18	&	14571.20		&	4437.14		&	-		&	{\bf 20.03}		&	36			&	13M			&	126			&	1,116		&	18		&	5.80 		\\
{\it audikw\_1}		&	36	&	*			&	9282.49		&	-		&	{\bf 190.17}	&	4,101		&	38M			&	59K			&	721K		&	36		&	58.38 	\\
{\it bone010}		&	24	&	*			&	10002.67		&	-		&	{\bf 393.11}	&	37K			&	34M			&	361K		&	44M			&	24		&	24.39 	\\ 
{\it af\_shell10}		&	15	&	*			&	21669.96		&	-		&	{\bf 50.99}		&	19			&	25M			&	75			&	2,105		&	15		&	10.67 	\\
{\it as-Skitter}		&	67	&	24385.73		&	*			&	-		&	{\bf 3838.36}	&	1M			&	6M			&	981K		&	737M		&	66		&	27.08 	\\ 
{\it roadNet-CA}	&	4	&	*			&	*			&	-		&	{\bf 0.44}		&	1M			&	1M			&	370K		&	4,302		&	4		&	0.08 		\\ 
{\it kkt\_power}		&	11	&	*			&	*			&	-		&	{\bf 2.26}		&	1M			&	4M			&	401K		&	2M			&	11		&	1.83 		\\ 
\midrule
{\it rmat\_er\_1}		&	3	&	256.37		&	215.18		&	49.79	&	{\bf 0.38}		&	780			&	1M			&	915			&	8,722		&	3		&	0.12 		\\
{\it rmat\_er\_2}		&	3	&	1016.70		&	865.18		&	-		&	{\bf 0.78}		&	2,019		&	2M			&	2,351		&	23K			&	3		&	0.24 		\\
{\it rmat\_er\_3}		&	3	&	4117.35		&	3456.39		&	-		&	{\bf 1.87}		&	4,349		&	4M			&	4,960		&	50K			&	3		&	0.49 		\\
{\it rmat\_er\_4}		&	3	&	16419.80		&	13894.52		&	-		&	{\bf 4.16}		&	9,032		&	8M			&	10K			&	106K		&	3		&	1.44 		\\
{\it rmat\_er\_5}		&	3	&	*			&	*			&	-		&	{\bf 9.87}		&	18K			&	16M			&	20K			&	212K		&	3		&	2.57 		\\
\midrule
{\it rmat\_sd1\_1}	&	6	&	225.93		&	214.99		&	50.08	&	{\bf 1.39}		&	39K			&	1M			&	23K			&	542K		&	6		&	0.45 		\\
{\it rmat\_sd1\_2}	&	6	&	912.44		&	858.80		&	-		&	{\bf 3.79}		&	90K			&	2M			&	56K			&	1M			&	6		&	0.98 		\\ 
{\it rmat\_sd1\_3}	&	6	&	3676.14		&	3446.02		&	-		&	{\bf 8.17}		&	176K		&	4M			&	106K		&	2M			&	6		&	1.78 		\\ 
{\it rmat\_sd1\_4}	&	6	&	14650.40		&	13923.93		&	-		&	{\bf 25.61}		&	369K		&	8M			&	214K		&	5M			&	6		&	4.05 		\\ 
{\it rmat\_sd1\_5}	&	6	&	*			&	*			&	-		&	{\bf 46.89}		&	777K		&	16M			&	455K		&	12M			&	6		&	9.39 		\\ 
{\it rmat\_sd2\_1}	&	26	&	427.41		&	213.23		&	{\bf 48.17}	&	242.20		&	110K		&	853K		&	88K			&	614M		&	26		&	32.83 	\\ 
{\it rmat\_sd2\_2}	&	35	&	4663.62		&	{\bf 851.84}	&	-		&	3936.55		&	232K		&	1M			&	195K		&	1B			&	35		&	95.89 	\\ 
{\it rmat\_sd2\_3}	&	39	&	13626.23		&	{\bf 3411.14}	&	-		&	10647.84		&	470K		&	3M			&	405K		&	1B			&	37		&	245.51 	\\ 
{\it rmat\_sd2\_4}	&	43	&	*			&	{\bf 13709.52}	&	-		&	*			&	*			&	*			&	*			&	*			&	42		&	700.05 	\\
{\it rmat\_sd2\_5}	&	N	&	*			&	*			&	-		&	*			&	*			&	*			&	*			&	*			&	51		&    1983.21 	\\
\midrule
{\it hamming6-4}	&	4	&	{\bf $<$0.01}	&	{\bf $<$0.01}	&{\bf $<$0.01}	&	{\bf$<$0.01}		&	0			&	704			&	0			&	0			&	4		&	$<$0.01 	\\
{\it johnson8-4-4}	&	14	&	0.19			&	{\bf $<$0.01}	&{\bf $<$0.01}	&	0.23			&	0			&	1,855			&	0			&	0			&	14		&	$<$0.01 	\\
{\it keller4}		&	11	&	22.19		&	0.15			&	{\bf 0.02}	&	23.35		&	0			&	9,435			&	0			&	0			&	11		&	$<$0.01 	\\
{\it c-fat200-5}		&	58	&	0.60			&	0.33			&	{\bf 0.01}	&	0.93			&	0			&	8,473			&	0			&	0			&	58		&	0.04 		\\
{\it brock200\_2}	&	12	&	0.98			&	0.02			&{\bf $<$0.01}	&	1.10			&	0			&	9,876			&	0			&	0			&	10		&	$<$0.01 	\\
\bottomrule\bottomrule
\end{tabular}

\end{table}

%% file: new_timings.tex
\begin{table}[tbh]
\centering
\caption{Comparison of runtimes of algorithms: \cite{ostergard} ({\it $\tau_{cliquer}$}), \cite{konc2007improved} ({\it $\tau_{MCQD+CS}$}), 
with that of our new exact algorithm ($\tau_{A1}$) for selected medium and large Pajek and Stanford data sets. 
An asterisk (*) indicates that the algorithm did not terminate within 
15,000 seconds for that instance. The column $\omega$ lists the maximum clique size in each graph, 
the column $\omega_{A_2}$ lists the clique size returned by our heuristic and the column
$\tau_{A_2}$ lists the heuristic's runtime.}
\label{tab:pajek_stanford}
\begin{tabular}{lr|rrr|rr}
\toprule\toprule
	&		&			&	$\tau_{MCQD}$	&		&		&		\\
$G$	&	$\omega$	&	$\tau_{cliquer}$	&	$_{+CS}$	&	$\tau_{A1}$	&	$\omega_{A2}$	&	$\tau_{A2}$	\\ \hline \hline
{\it foldoc}	&	9	&	2.23	&	0.58	&	0.05	&	9	&	$<$0.01	\\
{\it eatRS}	&	9	&	7.53	&	1.86	&	1.81	&	9	&	1.80	\\
{\it hep-th}	&	23	&	9.43	&	2.43	&	4.48	&	23	&	0.06	\\
{\it patents}	&	6	&	2829.48	&	239.66	&	0.13	&	6	&	0.03	\\
{\it days-all}	&	28	&	2.47	&	0.59	&	75.37	&	21	&	0.05	\\
\midrule
{\it roadNet-TX}	&	4	&	*	&	*	&	0.26	&	4	&	0.04	\\
{\it amazon0601}	&	11	&	3190.11	&	717.99	&	0.20	&	11	&	0.30	\\
{\it email-EuAll}	&	16	&	943.74	&	270.67	&	0.92	&	14	&	0.06	\\
{\it web-Google}	&	44	&	10958.68	&	*	&	0.35	&	44	&	0.6	\\
{\it soc-wiki-Vote}	&	17	&	0.89	&	0.24	&	4.31	&	14	&	0.02	\\
{\it soc-slashdot0902}	&	27	&	88.38	&	23.63	&	25.54	&	22	&	0.06	\\
{\it cit-Patents}	&	11	&	2.47	&	*	&	19.99	&	10	&	4.15	\\
{\it soc-Epinions1}	&	23	&	73.85	&	19.95	&	15.01	&	19	&	0.06	\\
{\it soc-wiki-Talk}	&	26	&	*	&	*	&	6885.13	&	18	&	0.45	\\
{\it web-berkstan}	&	201	&	6416.61	&	*	&	44.70	&	201	&	32.03	\\
	\bottomrule\bottomrule
\end{tabular}
\end{table}

%% file: parallelization.tex
\section{Parallelization}
\label{sec:parallelization}


 We demonstrate in this section how our exact algorithm can be parallelized and show performance results
on a shared-memory platform. The heuristic can be parallelized following a similar procedure.

As explained in Section \ref{sec:algorithms}, the $i$th iteration of the {\em for} loop in the exact algorithm (Algorithm 1) computes the size of the largest clique that contains the vertex $v_i$. Since our algorithm does not impose any specific order in which vertices have to be processed, these iterations can in principle be performed concurrently. During such a concurrent computation, however, different processes might discover maximum cliques of different sizes---and for the pruning steps to be most effective, the current globally largest maximum clique size needs to be communicated to all processes as soon as it is discovered. In a shared-memory programming model, the global maximum clique size can be stored as a shared variable accessible to all the processing units, and its value can be updated by the relevant processor at any given time. In a distributed-memory setting, more care needs to be exercised to keep the communication cost low. 

\begin{figure}[!h]
  \centering
    \includegraphics[scale=0.21]{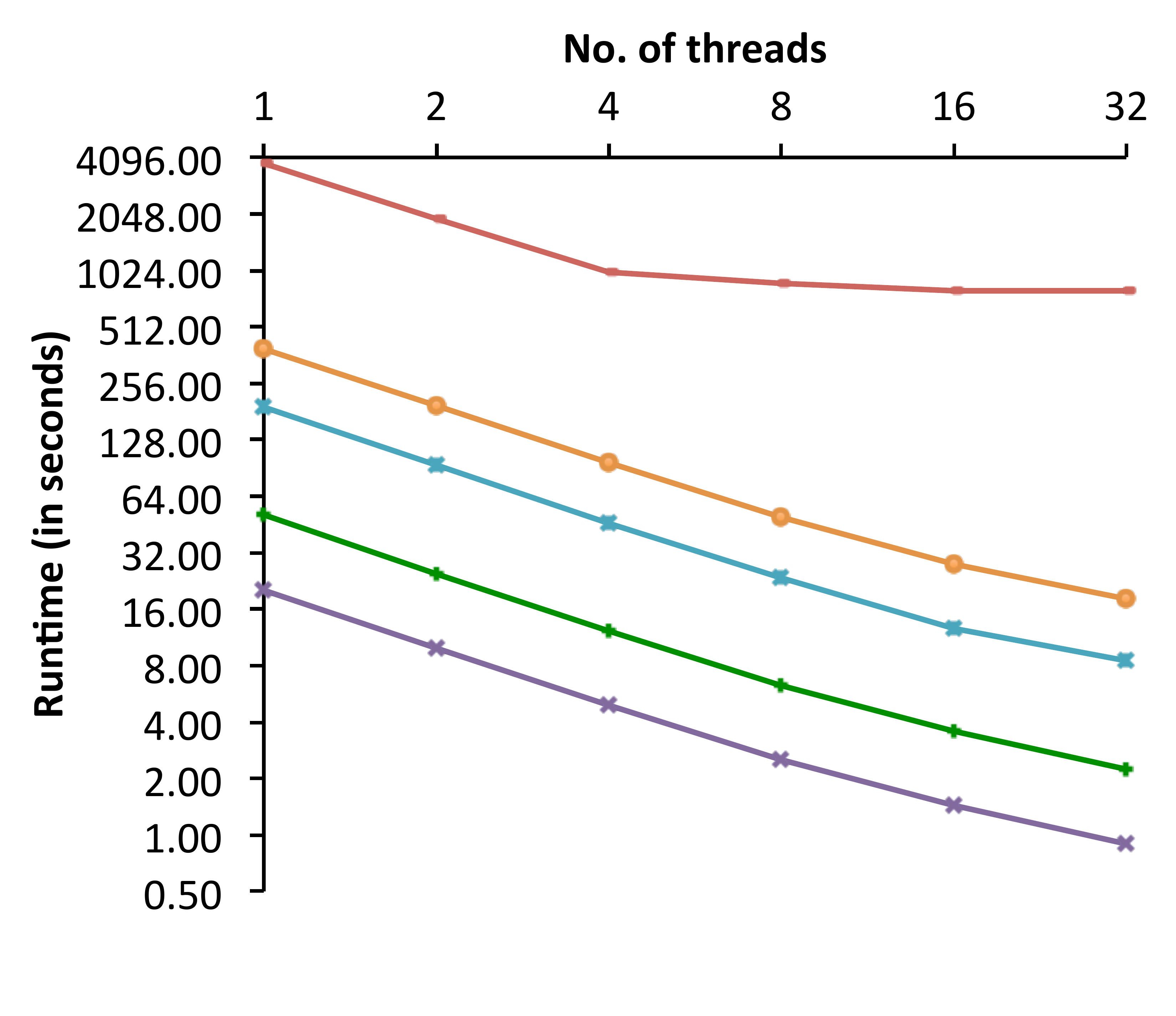}
    \includegraphics[scale=0.21]{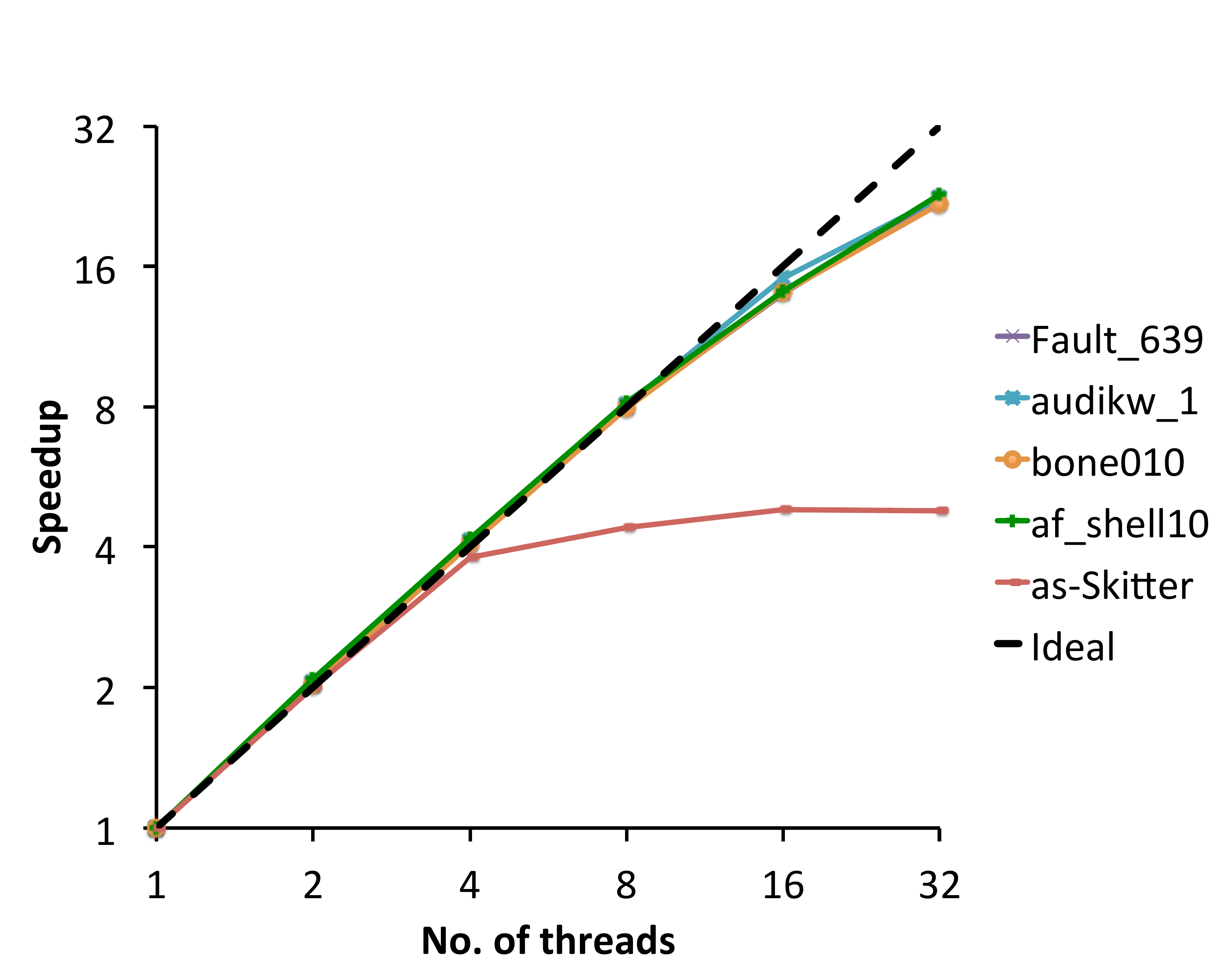}
    \includegraphics[scale=0.21]{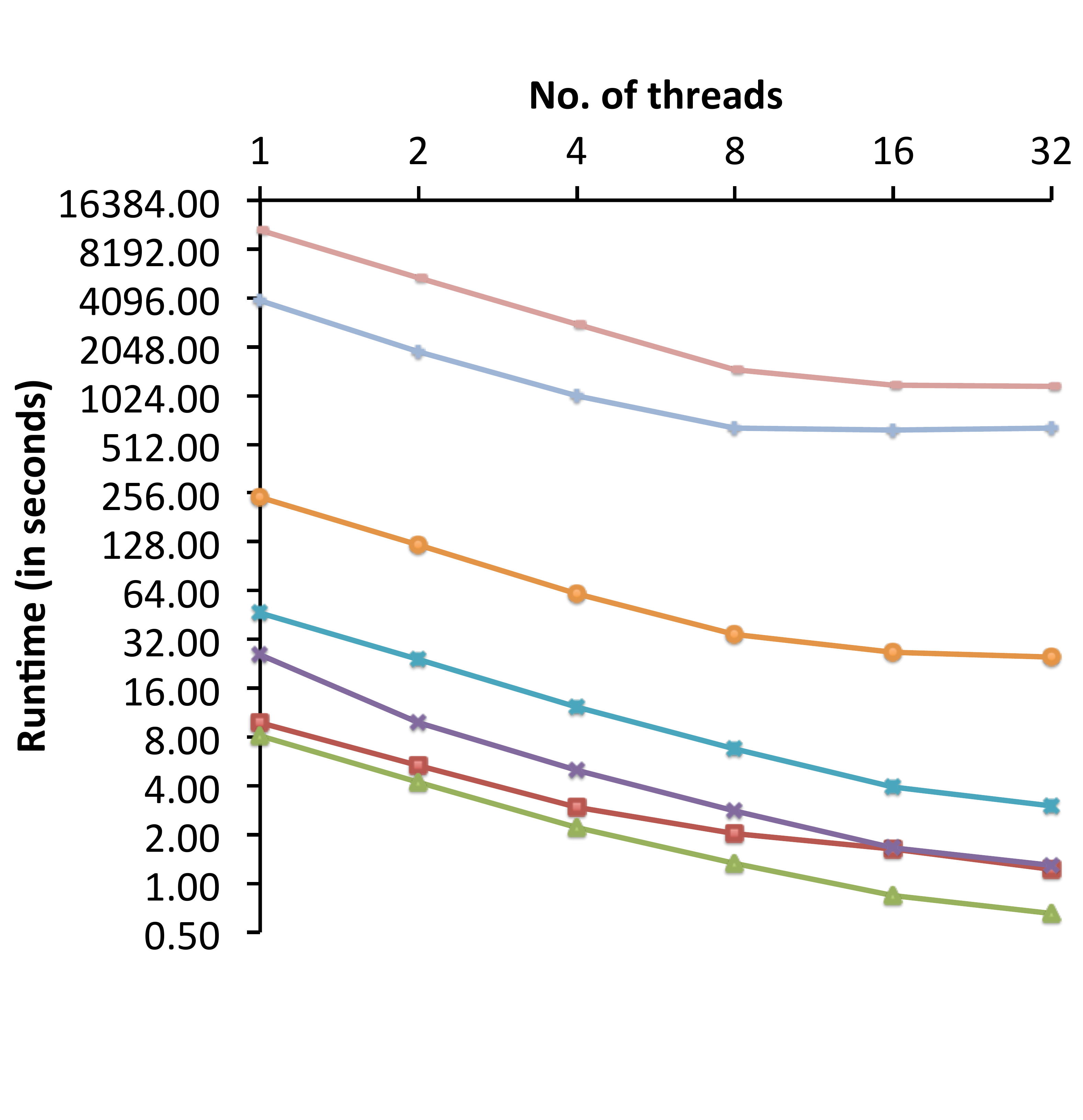}
    \includegraphics[scale=0.21]{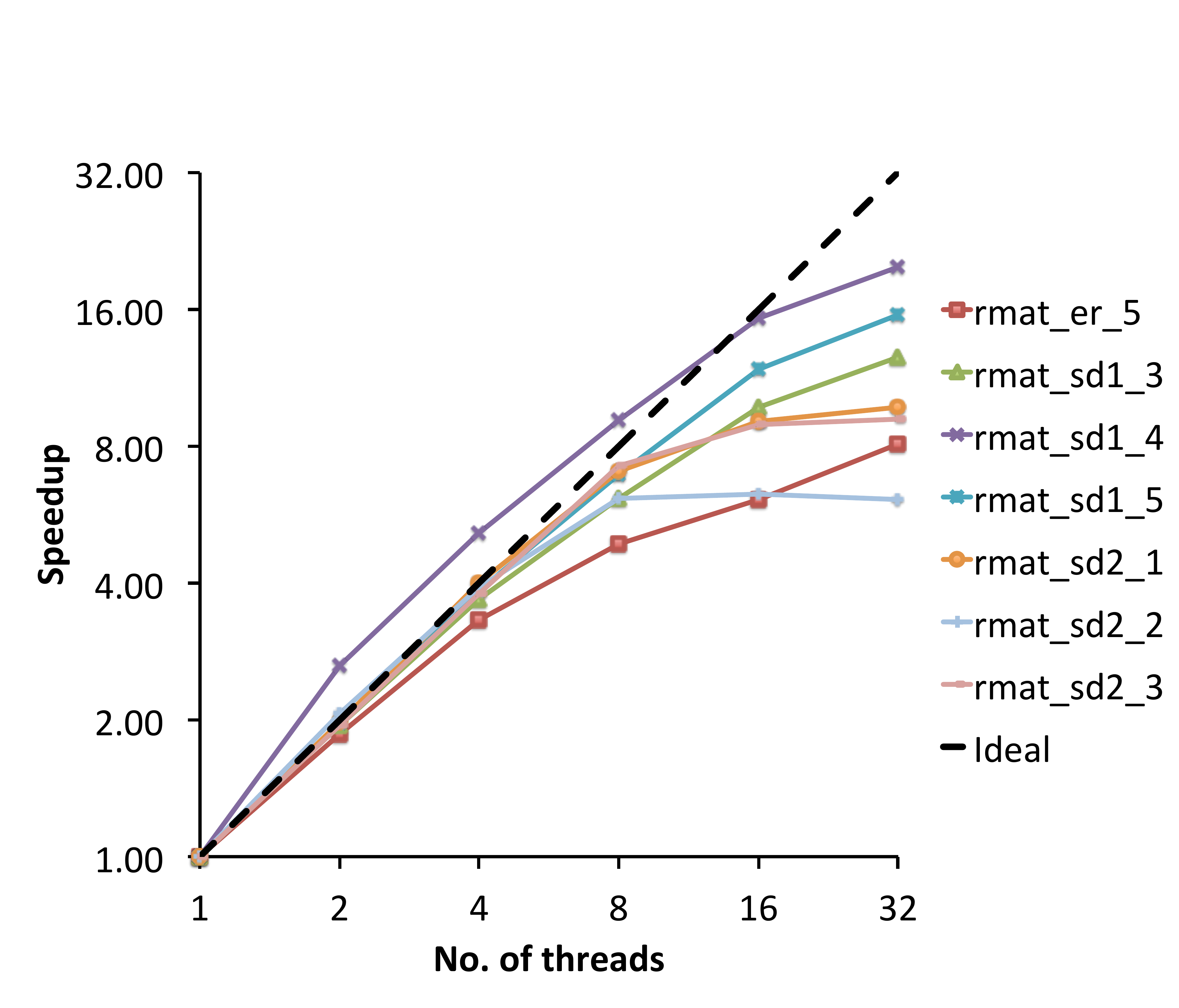}
    
 \caption{Performance (timing and speedup plots) of shared-memory parallelization on graphs in the test-bed. The top set of figures show performance on the real-world graphs, whereas the bottom set show results for the RMAT graphs. Graphs whose sequential runtime is less than 5 seconds were omitted.}
\label{fig-parallel_perf}
\end{figure}

We implemented a shared-memory parallelization based on the procedure described above using OpenMP. Since the global value of the maximum clique is shared by all processing units, we embed the step that updates the maximum clique, i.e. Line \ref{critical} of Algorithm \ref{alg:mClq}, into a {\it critical} section (an OpenMP feature that enforces a lock, thus allowing only one thread at a time to make the update). Further, we use dynamic load scheduling, since different vertices might return different sizes of maximum clique resulting in different work loads.

We performed experiments on the same graphs listed in the testbed described in Section \ref{sec:experiments}. For these, we used a Linux workstation 
with six 2.00GHz Intel Xeon E7540 processors. Each processor has six cores, each with 32KB of L1 and 256KB of L2 cache, and each processor shares an 18MB L3 cache.  

Figure \ref{fig-parallel_perf} shows the timings and speedups we obtained for the real-world and RMAT graphs separately. We omitted graphs whose sequential runtimes were less than 5 seconds as they were too low to measure and do meaningful assessment of parallelization performance. For most of the real-world graphs, one can see from the figure that we obtained near-linear scaling of runtimes and speedups when up to 16 threads are used. The only exception is the graph 
{\emph as-Skitter}. The relatively poorer scaling there is likely due to the fact that the instance has a relatively large maximum clique, and hence the core (thread) which computes it spends relatively large amount of time computing it while other cores (threads) which have completed their processing of the remaining vertices remain idle. 
For the other instances of the real-world graphs, the maximum speedup we obtained 
while using 32 cores/threads is around 22$\times$. 

For the RMAT graphs, it can be seen that the scaling of the runtime and speedups vary with the structures (RMAT parameters) of the instances. We observed super-linear speedups for a couple of the instances, which
happens as a result of some unfruitful searches in the branch-and-bound procedure being discovered early
as a result of parallel processing. This phenomenon is better exploited and more fully explored in other works in the literature such as  \cite{a6040618}.
For the other instances, the algorithm scales fairly well up to 8 threads, and begins to
degrade thereafter. The speedups we obtained range between 6$\times$  to 20$\times$ when 32 threads are used.

%% file: applications2.tex
\section{Clique Algorithms and Community Detection}
\label{sec:applications}

In this section we demonstrate how clique finding algorithms can be used for detecting overlapping communities in networks.

{\bf Background. }
Most community detection algorithms are designed to identify mutually independent communities in a given network and therefore are not suitable for detecting overlapping communities. Yet, in many real-world networks, it is natural to find vertices (or members) that belong to more than one group (or community) at the same time.

Palla et al.~\cite{cite-key} introduced the Clique Percolation Method (CPM) 
as one effective approach for detecting overlapping communities in a network. 
The basic premise in CPM is that a typical community is likely to be made up of several cliques that share many of their vertices.  
We recall a few notions defined in \cite{cite-key} to make this more precise.
A clique of size $k$ is called a {\em $k$-clique}, and   
two $k$-cliques are called {\em adjacent} if they share $k - 1$ nodes. 
A {\em $k$-clique community} is a union of all $k$-cliques that can be reached 
from each other through a series of adjacent $k$-cliques. 
With these notions at hand, Palla et al. devise a method to extract such $k$-clique communities of a network. 
Note that, by definition, $k$-clique communities allow for overlaps, i.e. common vertices can be shared by the  communities.

\begin{figure}
  \centering
    \includegraphics[scale=0.45]{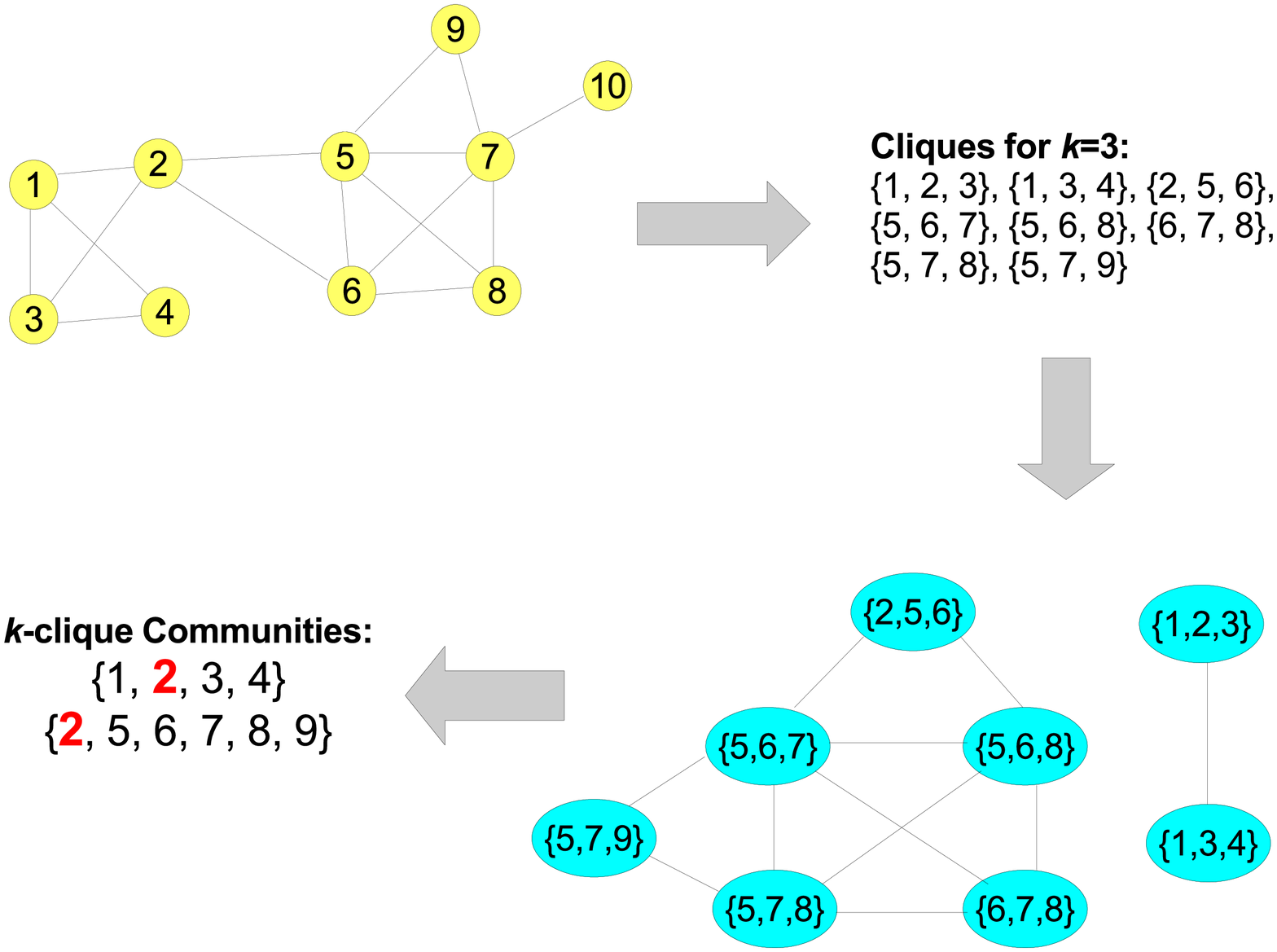}
    
  \caption{Illustration of overlapping community detection by the Clique Percolation Method (CPM) algorithm on a sample graph. Steps involved are 1) Detecting the $k$-cliques 2) Forming the clique-graph 3) Merging members of the connected components in the clique-graph to obtain the $k$-clique communities. In this example, node 2 is shared by the two communities formed, resulting in an overlapping structure.}
\label{fig-cpm}
\end{figure}

The CPM algorithm is illustrated in Figure \ref{fig-cpm}. Given a graph, we first extract all cliques of size $k$; for this example we choose $k$=3. This is followed by generating the {\it clique graph}, where each $k$-clique in the original graph is represented by a vertex. An edge is added between any two $k$-cliques in the clique graph that are adjacent. For the case $k$=3, this means an edge is added between any two 3-cliques in the clique graph that share two common vertices (of the original graph). The connected components in the clique graph represent a community, and the actual members of the community are obtained by gathering the vertices of the individual cliques that form the connected component. In our example in Figure \ref{fig-cpm}, we obtain two communities, which share a common vertex (vertex 2), forming an overlapping community structure.

A large clique of size $q \geq k $ contains $\binom{q}{k}$ different {\em k-cliques}.
An algorithm that tries to locate the $k$-cliques individually and examine the adjacency
between them can therefore be very very slow for large networks. 
Palla et al. make two observations that help one come up with a better strategy.
First, a clique of size $q$ is clearly a $k$-clique connected subset for any $k \leq q$. 
Second, two large cliques that share at least $k-1$ nodes form one $k$-clique connected component as well.
Leveraging these, the strategy of Palla et al. avoids searching for $k$-cliques individually
and instead first locates the large cliques in the network and then looks for the $k$-clique connected subsets of given $k$ (that is, the $k$-clique communities) by studying the overlap between them.
More specifically, their algorithm first constructs a symmetric {\em clique-clique overlap matrix}, 
in which each row represents a large (to be precise a maximal) clique, 
each matrix entry is equal to the number of common nodes between the two corresponding cliques, and each diagonal entry is equal to the size of the clique. 
The $k$-clique communities for a given $k$ are then equivalent to connected 
clique components in which the neighboring cliques are linked to each other by at least $k-1$ common nodes. Palla et al. then find these components by erasing every off-diagonal entry smaller than $k-1$ and every diagonal entry smaller than $k$ in the matrix, replacing the remaining elements by one, and then carrying out a component analysis of this matrix. The resulting separate components are equivalent to the different $k$-clique communities.



{\bf Our Method. } We devised an algorithm based on a similar idea as the procedure
above, but using a variant of our heuristic maximum clique-algorithm 
(Algorithm 2) for the core clique detection step. 

The standard CPM in essence presupposes finding {\em all} maximal cliques in a network.
The number of maximal cliques in a network can in general be exponential in the number of nodes $n$ in the network. In our method, we work instead with a smaller set of maximal cliques. In particular, the basic variant of our method considers exactly one clique per vertex, the largest of all the maximal cliques that it belongs to. Clearly, this can be too restrictive a requirement and may fail to add deserving members to a $k$-clique community. 
To allow for more refined solutions, we include a parameter $c$
in the method that tells us how many additional cliques per vertex would be considered.
The case $c=0$ corresponds to no additional cliques than the largest maximal clique containing the vertex. The case $c\geq1$ corresponds to $c$ additional maximal cliques (each of size at least $k$)
per vertex. Hence, in total, $n(c+1)$ cliques will be collected. For a particular vertex $v$, the largest maximal clique containing it 
can be heuristically obtained by exploring the neighbor of $v$ with {\em maximum degree} 
in the corresponding step in Algorithm 2.
In contrast, to obtain the other maximal cliques involving $v$ (case $c\geq1$),
we explore a {\em randomly} chosen neighbor of $v$, regardless of its degree value.

{\bf Test on Synthetic Networks. } We tested our algorithm on the LFR benchmarks\footnote{http://sites.google.com/site/andrealancichinetti/files} proposed in \cite{2008PhRvE..78d6110L}. These benchmarks have the attractive feature that they allow
for generation of synthetic networks with known communities (ground truth).
We generated graphs with $n$ = 1000 nodes, average degree $K$ = 10, and power law exponents $\tau_1$ = 2 and $\tau_2$ = 1 in the LFR model. We set the maximum degree $K_{max}$ to 50, and the minimum and maximum community sizes $C_{min}$ and $C_{max}$ to 20 and 50, respectively. We used two far-apart values each for the mixing parameter $\mu$, the fraction of overlapping nodes $O_n$, and the number of communities each overlapping node belongs to $O_m$. Specifically, for $\mu$ we used 0.1 and 0.3, for $O_n$ we used 10$\%$ and 50$\%$, and for $O_m$ we used 2 and 8. 
All together, these combinations of parameters resulted in eight graphs in the testbed.

We evaluated the performance of our algorithm against the ground truth 
using {\it Omega Index} \cite{doi:10.1207/s15327906mbr2302_6}, 
which is the overlapping version of the Adjusted Rand Index (ARI) \cite{ARI_paper_1985}.
Intuitively, Omega Index measures the extent of agreement between two given sets of communities, by looking at node pairs that occur the same number of times (possibly none) in both.
We used three values for the parameter $c$ of our method: 0, 2, and 5. 
Recall that $c$ set to zero corresponds to picking only the largest clique for each node. 
As $c$ is increased, more and more large cliques (each of size at least $k$)
are considered for each node. 
We compare the performance of our method with that of CFinder\footnote{http://www.cfinder.org}, an implementation of CPM~\cite{cite-key}. We used the command line utility provided in the package for all experiments. For this study, we used a MacBook Pro running OS X v.10.9.2 with 2.66GHz Intel Core i7 processor with 2 cores, 256KB of L2 cache per core, 4MB of L3 cache and 8GB of main memory.

\input{results_application1}


Table~\ref{tab:applications1} shows the experimental results we obtained.
The first three columns of the Table list the various parameters used to generate each test network. The fourth column lists the number of ground truth communities ($C(GT)$) in each network.  The remaining columns in the table show performance 
in terms of the number of communities detected ($C$), 
total number of shared nodes ($S$) and the Omega Index ($\Omega$) for 
CFinder and our method with different values for the parameter $c$.
In our experiments, for CFinder as well as all variants of our algorithm, 
we found that the value of $k=4$ gives the best Omega Index value relative 
to the ground truth. All results reported in Table~\ref{tab:applications1} are therefore
for $k=4$.

In general we observe that there is a close correlation between our method and CFinder
in terms of all three of the quantities $C$, $S$ and $\Omega$. It can be seen that as we increase the value of $c$ in our method, the Omega Index values get closer and closer to that of CFinder. For our algorithm run with $c=0$, the Omega Index is about 75$\%$ of that of CFinder. When run with $c$ = 2, it is about 92$\%$ and for $c$ = 5, it is about 99$\%$. When we increased the value of $c$ even further to 10, we observed that the Omega Index was almost identical to that obtained by CFinder. From this, one can see that we can get almost exactly the same results as the CPM method using our algorithm which uses only a small set of 
maximal cliques, as opposed to all the maximal cliques in the graph.

\input{results_application2}

Table~\ref{tab:applications2} shows the time taken by CFinder and our method 
run with $c$ = 0, 2 and 5. For our method, the table lists the total run time $\tau$ 
as well as the time $\tau_c$ spent on just the clique detection part and the ratio
${\tau_c}/{\tau}$ expressed in percents ($\tau_c$\%).
The remaining time $\tau - \tau_c$ is spent on building the clique-clique overlap matrix, eliminating cliques, component analysis and generating the communities.  As a side note,
we point out that our immediate goal in the implementation of these later phases has been quick experimentation, rather than efficient code, and therefore there is a very large room for improving the runtimes. This is also reflected by the numbers; one can see from the table that the average percentage of time our algorithm spends on clique-finding decreases as $c$ is increased; the quantity is about 30$\%$ when $c$ = 0 and about 16$\%$ when $c$ = 5. Yet, looking only at the total time taken in Table~\ref{tab:applications2}, and comparing CFinder and our algorithm run with $c$ = 5 (the case where the Omega Index values match most closely with CFinder), we see that our algorithm is at least 4$\times$ faster on average. When $c$ is set to 0 and 2, the mean speedups are 51$\times$ and 13$\times$ respectively. 

Since the source code for CFinder is not publicly available, we were not able to measure exactly the proportion of time it spends on clique-finding. 
We suspect it constitutes a vast proportion of the total runtime.


\input{results_application3}

{\bf Test on Real-world Networks. } We also tested our method and CFinder on four of the real world graphs from our testbed in Section \ref{sec:experiments}---{\em cond-mat-2003}, {\em email-Enron}, {\em dictionary28} and {\em roadNet-CA}---and a user-interest-based graph generated using data collected from Facebook. We briefly explain here how this graph was generated.
Every user on Facebook has a {\it wall}, which is a the user's profile space that allows the posting of messages, often short or temporal notes or comments by other users. We generate a graph with the {\it walls} as vertices, and assign an edge between a pair of vertices, if there is at least one user who has commented on both walls. We assign edge weights proportional to the number of common users, as we consider this to be an indicator of the strength of the connection. A more elaborate explanation of the data collection method and network generation for the Facebook graph is discussed in \cite{dianapaper}. As this is a weighted graph, we used a threshold (of 0.009) to include only strong links, resulting in a network with 1144 vertices and 2561 edges.

\begin{figure}[h!]
  \centering
    \includegraphics[scale=0.4]{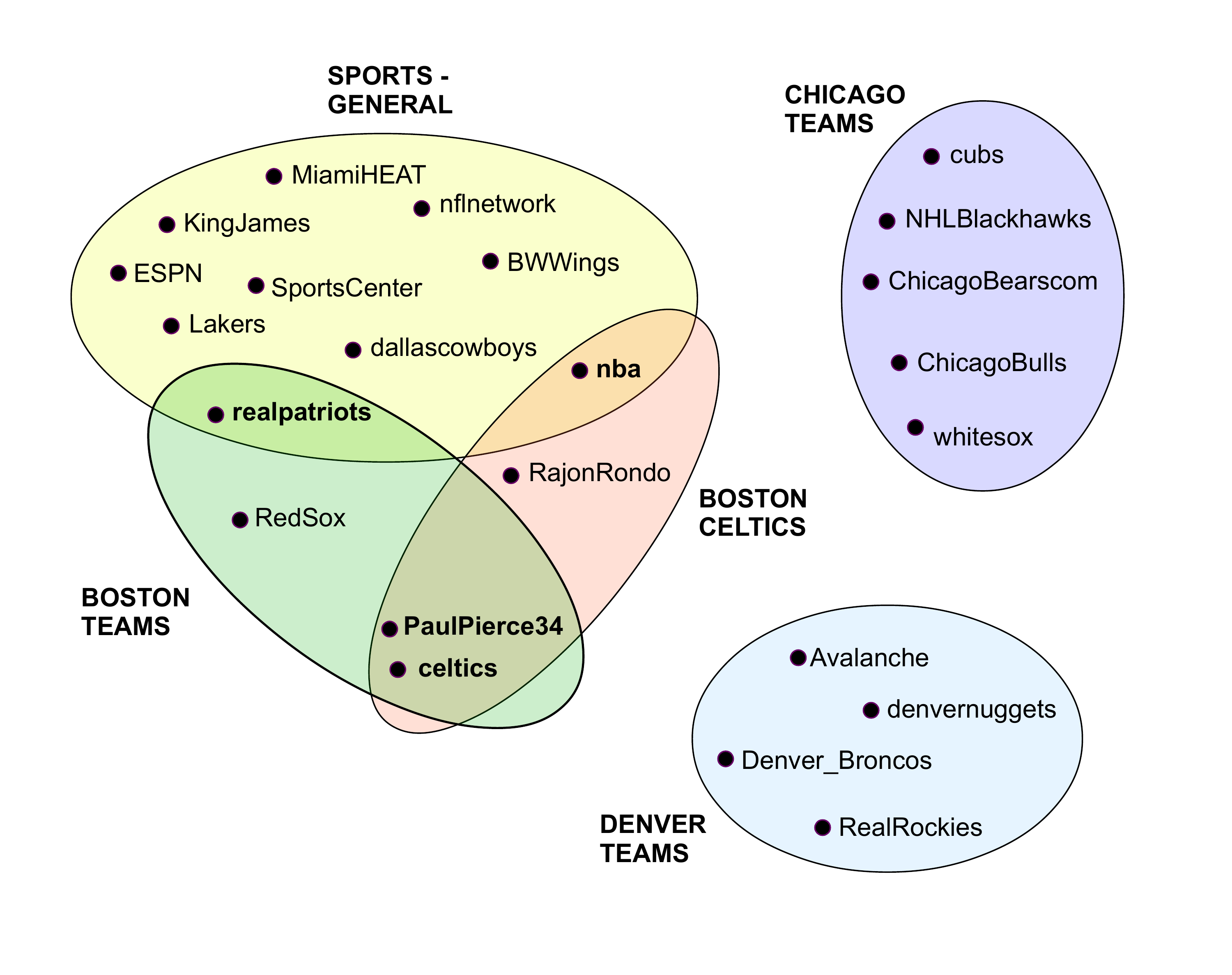}
    
\vspace{-15pt}
  \caption{Some Facebook communities detected by our algorithm.}
\label{fig-communities-fb}
\end{figure}

The results of our method for $c$ = 0 and 5, $k$ = 4, and CFinder for the case $k$ = 4 on all the five real-world graphs is shown in  Table~\ref{tab:applications3}. The table lists the number of communities found by each algorithm, total number of shared nodes, and the total time taken. For our algorithm, it also specifies the time spent on clique-finding. These graphs do not have any known community structure, so we were not able to measure the Omega Index values. For the graph {\em email-Enron}, CFinder was not able complete within 30 minutes. For the $c$ = 0 case, our algorithm comfortably outperforms CFinder, even in terms of total time, on average performing about 115$\times$ faster. For the $c$ = 5 case, the total time taken by our algorithm turns out to be higher for some cases, and lower for others. However, it should be noted that the time taken for clique finding is incomparably smaller than the post-clique-finding processing time. 

We also present some of the sports-related communities detected by our algorithm in the Facebook graph in Figure \ref{fig-communities-fb}, which is self-explanatory.

%% file: results_application1.tex
\begin{table}[!h]

\footnotesize
\small
\centering
\caption{Results of communities detected using our new method and CFinder on LFR benchmarks \cite{2008PhRvE..78d6110L}. 
All networks have $n$ = 1000 nodes, and the parameters used to generate the netwroks are listed in the first three columns: $\mu$ is the mixing parameter, $O_m$ is the number of communities each overlapping node is assigned to $O_n$ is the fraction of overlapping nodes. $C(GT)$ is the number of communities in the input graph (ground truth). $C$ denotes the number of communities, $S$ the number of shared nodes, and $\Omega$ the omega index. }
\label{tab:applications1}

\begin{tabular}{ c c c c || c@{\hspace{3pt}} c@{\hspace{3pt}} c@{\hspace{3pt}} | c c c | c c c | c c c }
\toprule\toprule
	&  			& 			& 			&  		& 	CFinder	&	&	\multicolumn{9}{|c}{Our Method}			\\
	&  			& 			& 			&  		& 		&		&	\multicolumn{3}{c}{$c=0$}&	\multicolumn{3}{c}{$c=2$}	&		\multicolumn{3}{c}{$c=5$}		\\
\hline
$\mu$&	$O_m$	&	$O_n$	&	$C(GT)$	&	$C$	&	$S$	&	$\Omega$	&	$C$	&	$S$		&	$\Omega$	&	$C$	&	S		&	$\Omega$	&	$C$	&	$S$	&	$\Omega$	\\
\hline\hline
0.1	&	2	&	10\%	&	34	&	37	&	55	&	0.842	&	60	&	64	&	0.758	&	49	&	76	&	0.825	&	40		&	59	&	0.840	\\
0.1	&	2	&	50\%	&	44	&	62	&	153	&	0.402	&	87	&	116	&	0.280	&	83	&	158	&	0.363	&	68		&	158	&	0.395	\\
0.1	&	8	&	10\%	&	51	&	58	&	30	&	$\sim$0.0	&	76	&	46	&	$\sim$0.0	&	65	&	39	&	$\sim$0.0	&	61		&	35	&	$\sim$0.0	\\
0.1	&	8	&	50\%	&	130	&	74	&	68	&	0.030	&	74	&	56	&	0.020	&	79	&	66	&	0.028	&	79		&	72	&	0.030	\\
0.3	&	2	&	10\%	&	35	&	50	&	53	&	0.612	&	69	&	67	&	0.461	&	64	&	73	&	0.577	&	55		&	60	&	0.605	\\
0.3	&	2	&	50\%	&	44	&	60	&	95	&	0.193	&	92	&	93	&	0.109	&	85	&	102	&	0.152	&	77		&	106	&	0.169	\\
0.3	&	8	&	10\%	&	48	&	64	&	35	&	0.304	&	79	&	59	&	0.239	&	74	&	55	&	0.293	&	69		&	46	&	0.301	\\
0.3	&	8	&	50\%	&	142	&	57	&	43	&	0.019	&	58	&	46	&	0.013	&	61	&	49	&	0.017	&	59		&	45	&	0.018	\\
\bottomrule\bottomrule
\end{tabular}

\end{table}

%% file: results_application2.tex
\begin{table}[!h]

\footnotesize
\small
\centering
\caption{Timing results of our new method and CFinder on the LFR benchmarks. All networks have $n$ = 1000 nodes, and the parameters used to generate the graphs are given by the first three columns: $\mu$ is the mixing parameter,  $O_m$ is the number of communities each overlapping node is assigned to and $O_n$ is
the fraction of overlapping nodes, $\tau_c$ is the time spent (in seconds) on computing the cliques, $\tau$ is the total time (in seconds), and $\tau_c\%$ is the percentage of time spent for clique computation. }
\label{tab:applications2}

\begin{tabular}{ c c c || c | c c c | c c c | c c c }
\toprule\toprule
 	&  		& 		&	CFinder	&			\multicolumn{9}{c}{Our Method}			\\
 	&  		& 		&			&			\multicolumn{3}{c}{$c=0$}				&		\multicolumn{3}{c}{$c=2$}	&	\multicolumn{3}{c}{$c=5$}				\\
\hline
$\mu$&$O_m$	&$O_n$	&	$\tau$	&	$\tau_c$	&	$\tau$	&$\tau_c\%$	&	$\tau_c$	&	$\tau$	&$\tau_c\%$	&	$\tau_c$	&	$\tau$	&$\tau_c\%$	\\
\hline\hline
0.1	&	2	&	10\%	&	0.913	&	0.005	&	0.031	&	17.3\%	&	0.025	&	0.255	&	9.7\%	&	0.050	&	0.912	&	5.5\%	\\
0.1	&	2	&	50\%	&	0.855	&	0.004	&	0.016	&	22.0\%	&	0.010	&	0.089	&	10.8\%	&	0.017	&	0.254	&	6.8\%	\\
0.1	&	8	&	10\%	&	0.732	&	0.004	&	0.025	&	17.4\%	&	0.012	&	0.159	&	7.4\%	&	0.024	&	0.550	&	4.4\%	\\
0.1	&	8	&	50\%	&	0.358	&	0.002	&	0.005	&	42.8\%	&	0.006	&	0.016	&	36.1\%	&	0.011	&	0.042	&	26.7\%	\\
0.3	&	2	&	10\%	&	0.592	&	0.004	&	0.013	&	26.0\%	&	0.011	&	0.090	&	11.8\%	&	0.027	&	0.321	&	8.3\%	\\
0.3	&	2	&	50\%	&	0.434	&	0.002	&	0.006	&	38.1\%	&	0.006	&	0.024	&	27.2\%	&	0.013	&	0.068	&	18.4\%	\\
0.3	&	8	&	10\%	&	0.472	&	0.003	&	0.011	&	25.7\%	&	0.008	&	0.064	&	12.5\%	&	0.016	&	0.205	&	7.6\%	\\
0.3	&	8	&	50\%	&	0.326	&	0.002	&	0.005	&	50.0\%	&	0.006	&	0.011	&	54.9\%	&	0.017	&	0.034	&	51.3\%	\\\bottomrule\bottomrule
\end{tabular}

\end{table}

%% file: results_application3.tex
\begin{table}[!h]

\footnotesize
\small
\centering
\caption{Results of our new algorithm and CFinder on three real-world graphs from Table \ref{tab:struc-graphs} and one additional graph generated using data collected from Facebook. The structural properties of the former three graphs are listed in Table \ref{tab:struc-graphs}. The Facebook graph has 1144 vertices and 2561 edges after thresholding. $C$ is the total number of communities found, $S$ is the total number of shared nodes, 
$\tau_c$ is the time spent (in seconds) on computing the cliques, and $\tau$, the total time (in seconds). A '-' denotes the algorithm did not complete in 30 minutes.}
\label{tab:applications3}

\begin{tabular}{ l || c c c | c c c c | c c c c }
\toprule\toprule
 					&  		\multicolumn{3}{c}{CFinder}			&		\multicolumn{8}{|c}{Our Method}			\\
 					&  		\multicolumn{3}{c}{}			&		\multicolumn{4}{|c}{$c=0$}			&	\multicolumn{4}{c}{$c=5$}		\\
\hline
$Name$				&	$C$	&	S	&$\tau$		&	$C$	&	$S$	&	$\tau_c$	&	$\tau$	&	C	&	S	&$\tau_c$		&	$\tau$	\\
\hline\hline
{\em cond-mat-2003}		&	4132	&	5180	&	68.46	&	3081	&	3937	&	0.130	&	14.000	&	3070	&	4460	&	0.917	&	417.151	\\
{\em email-Enron}		&	-	&	-	&	-		&	3947	&	4229	&	0.096	&	7.505	&	2518	&	3899	&	0.819	&	353.236	\\
{\em dictionary28}		&	3675	&	4308	&	9.67		&	2185	&	1799	&	0.040	&	1.132	&	3222	&	4154	&	0.300	&	95.596	\\
{\em roadNet-CA}		&	41	&	0	&	312.21	&	42	&	0	&	1.788	&	2.290	&	42	&	0	&	13.287	&	21.620	\\
{\em facebook}	&	29	&	29	&	3.42		&	38	&	46	&	0.002	&	0.011	&	19	&	27	&	0.023	&	0.252	\\
\bottomrule\bottomrule
\end{tabular}

\end{table}

%% file: conclusion.tex
\section{Conclusion}
\label{sec:conclusion}

We presented a new exact and a new heuristic algorithm for the maximum clique problem.
We performed extensive experiments on three broad categories of graphs comparing the 
performance of our algorithms to the algorithms by
Carraghan and Pardalos (CP) \cite{pardalos},
\"{O}sterg{\aa}rd ({\it cliquer}) \cite{ostergard}, Tomita et al. (MCQ, MCS) \cite{citeulike:7905505}\cite{walcom}, Konc and Jane\v{z}i\v{c} (MCQD) \cite{konc2007improved}, and Segundo et al. (BBMC)~\cite{SanSegundo}.
For DIMACS benchmark graphs and certain dense synthetic graphs ({\it rmat\_sd2}), our new exact algorithm performs comparably with the CP algorithm, but slower than the others. 
For large sparse graphs, both synthetic and real-world, our new algorithm runs
several orders of magnitude faster than the others. 
And its general runtime is observed to grow nearly linearly with the size of the graphs. 
The heuristic, which runs orders of magnitude faster than our exact algorithm and the others, delivers an optimal solution for 83\% of the test cases, and when it is sub-optimal, its accuracy ranges between 0.83 and 0.99 for the graphs in our testbed.
We also showed how the algorithms can be parallelized. Finally as a demonstration to the applications of clique finding algorithms, we presented a novel and efficient alternative based on our algorithms to the Clique Percolation Method \cite{cite-key}, to detect overlapping communities in networks. 

Maximum clique detection is often avoided by practitioners from being used as a component in 
a network analysis algorithm on the grounds of its NP-hardness. The results shown here suggest that this is not necessarily true, as maximum cliques can in fact be detected rather quickly for most real-world networks that are characterized by sparsity and other structures well suited for branch-and-bound type algorithms.



%% file: appendix_3.tex
\section*{Appendix}
\label{sec:appendix}

\begin{table}[!hbt]
\centering
\caption{Comparison of runtimes of our new exact maximum clique finding algorithm ($\tau_{A1}$) and the maximal clique enumeration algorithm by Eppstein \& Strash \cite{sea} ($\tau_{ES}$) for the Pajek and Stanford data sets in our testbed. The runtimes under column $\tau_{ES}$ are directly quoted from \cite{sea}.
The column $\Delta$ denotes the maximum degree, and $\omega$ the maximum clique size of the graph.}
\label{tab:pajek_stanford-es}
\begin{tabular}{lrrrr|rr}
\toprule\toprule
	&		&		&		&		&		&		\\
$G$	&	$\left|V\right|$	&	$\left|E\right|$	&	$\Delta$	&	$\omega$	&	$\tau_{A1}$	&	$\tau_{ES}$ \cite{sea}  	\\ \hline \hline
{\it foldoc}	&	13,356	&	91,470	&	728	&	9	&	0.05	&	0.13	\\
{\it eatRS}	&	23,219	&	304,938	&	1090	&	9	&	1.81	&	1.03	\\
{\it hep-th}	&	27,240	&	341,923	&	2411	&	23	&	4.48	&	1.70	\\
{\it patents}	&	240,547	&	560,943	&	212	&	6	&	0.13	&	1.65	\\
{\it days-all}	&	13,308	&	148,035	&	2265	&	28	&	75.37	&	5.18	\\
\midrule
{\it roadNet-TX}	&	1,393,383	&	1,921,660	&	12	&	4	&	0.26	&	4.00	\\
{\it amazon0601}	&	403,394	&	2,247,318	&	2752	&	11	&	0.20	&	6.03	\\
{\it email-EuAll}	&	265,214	&	364,481	&	7636	&	16	&	0.92	&	1.33	\\
{\it web-Google}	&	916,428	&	4,322,051	&	6332	&	44	&	0.35	&	9.70	\\
{\it soc-wiki-Vote}	&	8,297	&	100,762	&	1065	&	17	&	4.31	&	1.14	\\
{\it soc-slashdot0902}	&	82,168	&	504,230	&	2252	&	27	&	25.54	&	2.58	\\
{\it cit-Patents}	&	3,774,768	&	16,518,947	&	793	&	11	&	19.99	&	58.64	\\
{\it soc-Epinions1}	&	75,888	&	405,740	&	3044	&	23	&	15.01	&	4.78	\\
{\it soc-wiki-Talk}	&	2,394,385	&	4,659,565	&	100029	&	26	&	6885.13	&	216.00	\\
{\it web-berkstan}	&	685,230	&	6,649,470	&	84230	&	201	&	44.70	&	20.87	\\
\bottomrule\bottomrule
\end{tabular}
\end{table}

%% file: appendix_2.tex
\begin{table}[!hbt]
\centering
\caption{$P1$, $P2$, $P3$, $P4$ and $P5$ are the number of vertices pruned in steps Pruning 1, 2, 3, 4, and 5 of Algorithm 1. An asterisk (*) indicates that the algorithm did not terminate within 25,000 seconds for that instance. $\omega$ denotes the maximum clique size. For some of the graphs, none of the algorithms computed the maximum clique size in a reasonable time; the entry for the maximum clique size is marked with N, standing for ``Not Known''}
\label{tab:prunings}
\begin{tabular}{l@{\hspace{6pt}}r@{\hspace{6pt}}|@{\hspace{6pt}}r@{\hspace{6pt}}r@{\hspace{6pt}}r@{\hspace{6pt}}r@{\hspace{6pt}}r}

\toprule\toprule
$G$ 				& $\omega$ 	& 	$P1$ 		&	$P2$		& 	$P3$ 		& 	$P4$ 		&	$P5$		\vspace{-4pt} \\
				& 			&		 		& 		 		&				& 			 	&				\\ \hline \hline
{\it cond-mat-2003} 	& 	25 		& 	29,407 		&	48,096		&	6,527 		&	2,600		& 	17,576		\\
{\it email-Enron} 	& 	20 		& 	32,462 		&	155,344		&	4,060 		&	110,168		& 	8,835,739		\\ 
{\it dictionary28} 	& 	26 		& 	52,139		& 	4,353		&	2,114		&	542			& 	107			\\	
{\it Fault\_639}		&	18		&	36			&	13,987,719	&	126			&	10,767,992	&	1,116		\\
{\it audikw\_1}		&	36		&	4,101		&	38,287,830	&	59,985		&	32,987,342	&	721,938		\\
{\it bone010}		&	24		&	37,887		&	34,934,616	&	361,170		&	96,622,580	&	43,991,787	\\ 
{\it af\_shell10}		&	15		&	19			&	25,582,015	&	75			&	40,629,688 &	2,105		\\
{\it as-Skitter}		&	67		&	1,656,570		&	6,880,534		&	981,810		& 26,809,527&	737,899,486	\\ 
{\it roadNet-CA}	&	4		&	1,487,640		&	1,079,025		&	370,206		&	320,118		&	4,302		\\ 
{\it kkt\_power}		&	11		&	1,166,311		&	4,510,661		&	401,129		&	1,067,824		&	1,978,595		\\ 
\midrule
{\it rmat\_er\_1}		&	3		&	780			&	1,047,599		&	915			&	118,461		&	8,722		\\
{\it rmat\_er\_2}		&	3		&	2,019		&	2,094,751		&	2,351		&	235,037		&	23,908		\\
{\it rmat\_er\_3}		&	3		&	4,349		&	4,189,290		&	4,960		&	468,086		&	50,741		\\
{\it rmat\_er\_4}		&	3		&	9,032		&	8,378,261		&	10,271		&	933,750		&	106,200		\\
{\it rmat\_er\_5}		&	3		&	18,155		&	16,756,493	&	20,622		&	1,865,415		&	212,838		\\
\midrule
{\it rmat\_sd1\_1}	&	6		&	39,281		&	1,004,660		&	23,898		&	151,838		&	542,245		\\
{\it rmat\_sd1\_2}	&	6		&	90,010		&	2,004,059		&	56,665		&	284,577		&	1,399,314		\\ 
{\it rmat\_sd1\_3}	&	6		&	176,583		&	4,013,151		&	106,543		&	483,436		&	2,677,437		\\ 
{\it rmat\_sd1\_4}	&	6		&	369,818		&	8,023,358		&	214,981		&	889,165		&	5,566,602		\\\ 
{\it rmat\_sd1\_5}	&	6		&	777,052		&	16,025,729	&	455,473		&	1,679,109		&	12,168,698	\\ 
{\it rmat\_sd2\_1}	&	26		&	110,951		&	853,116		&	88,424		&	1,067,824		&	614,813,037	\\ 
{\it rmat\_sd2\_2}	&	35		&	232,352		&	1,645,086		&	195,427		&			81,886,879	&	1,044,068,886	\\ 
{\it rmat\_sd2\_3}	&	39		&	470,302		&	3,257,233		&	405,856		&			45,841,352	&	1,343,563,239	\\ 
{\it rmat\_sd2\_4}	&	43		&	*			&	*			&	*			&		*		&	*			\\
{\it rmat\_sd2\_5}	&	N		&	*			&	*			&	*			&		*		&	*			\\
\midrule
{\it hamming6-4}	&	4		&	0			&	704			&	0			&	583			&	0			\\
{\it johnson8-4-4}	&	14		&	0			&	1855			&	0			&	136,007		&	0			\\
{\it keller4}		&	11		&	0			&	9435			&	0			&	8,834,190		&	0			\\
{\it c-fat200-5}		&	58		&	0			&	8473			&	0			&	70449		&	0			\\
{\it brock200\_2}	&	12		&	0			&	9876			&	0			&	349,427		&	0			\\
\bottomrule\bottomrule
\end{tabular}
\end{table}

%% file: appendix_1.tex
\begin{table}[tbh]
\footnotesize
\centering
\caption{Comparison of runtimes of algorithms: \cite{pardalos} ({\it CP}), 
\cite{ostergard} ({\it $\tau_{cliquer}$}), \cite{konc2007improved} ({\it $\tau_{MCQD+CS}$}), \cite{prosser2012,citeulike:7905505} ({\it $\tau_{MCQ1}$}), \cite{prosser2012,walcom} ({\it $\tau_{MCSa1}$}), and \cite{prosser2012,SanSegundo} ({\it $\tau_{BBMC1}$}).
with that of our new exact algorithm ($\tau_{A1}$) for DIMACS graphs. 
An asterisk (*) indicates that the algorithm did not terminate within 
7,200 seconds for that instance. $\omega$ denotes the maximum clique size, $\omega_{A2}$ the maximum clique size found by our heuristic and $\tau_{A2}$, its runtime. For some of the graphs, none of the algorithms computed the maximum clique size in a reasonable time; the entry for the maximum clique size is marked with N, for ``Not Known''.}

\label{tab:dimacs}
\begin{tabular}{l@{\hspace{6pt}}r@{\hspace{6pt}}r@{\hspace{6pt}}r@{\hspace{6pt}}|@{\hspace{6pt}}r@{\hspace{6pt}}r@{\hspace{6pt}}r@{\hspace{6pt}}r@{\hspace{6pt}}r@{\hspace{6pt}}r@{\hspace{6pt}}r@{\hspace{6pt}}|@{\hspace{6pt}}r@{\hspace{3pt}}r}
\toprule\toprule
				& 		&     			&   		&  		& 		&$\tau_{MCQD}$ & &  & & & & \\
$G$	&	$\left|V\right|$	&	$\left|E\right|$	&	$\omega$	&	$\tau_{CP}$	&	$\tau_{cliquer}$	&	$_{+CS}$	&	$\tau_{MCQ1}$	&	$\tau_{MCSa1}$	&	$\tau_{BBMC1}$	&	$\tau_{A1}$	&	$\omega_{A2}$	&	$\tau_{A2}$ \\	\hline \hline
{\it brock200\_1}	&	200	&	14,834	&	21	&	*	&	10.37	&	0.75	&	7.11	&	5.48	&	1.7	&	*	&	18	&	0.02	\\
{\it brock200\_2}	&	200	&	9,876	&	12	&	0.98	&	0.02	&	0.01	&	0.1	&	0.1	&	$<$0.01	&	1.1	&	10	&	$<$0.01	\\
{\it brock200\_3}	&	200	&	12,048	&	15	&	14.09	&	0.16	&	0.03	&	0.35	&	0.4	&	0.1	&	14.86	&	12	&	$<$0.01	\\
{\it brock200\_4}	&	200	&	13,089	&	17	&	60.25	&	0.7	&	0.12	&	0.88	&	1.11	&	0.2	&	65.78	&	14	&	$<$0.01	\\
{\it brock400\_1}	&	400	&	59,723	&	27	&	*	&	*	&	671.24	&	4145.68	&	2873.91	&	762.14	&	*	&	20	&	$<$0.01	\\
{\it brock400\_2}	&	400	&	59,786	&	29	&	*	&	*	&	272.31	&	2848.1	&	2123.73	&	546.84	&	*	&	20	&	$<$0.01	\\
{\it brock400\_3}	&	400	&	59,681	&	31	&	*	&	*	&	532.77	&	2186.3	&	1523.55	&	431.92	&	*	&	20	&	$<$0.01	\\
{\it brock400\_4}	&	400	&	59,765	&	33	&	*	&	*	&	266.43	&	1038.33	&	881.31	&	211.29	&	*	&	22	&	$<$0.01	\\
{\it brock800\_1}	&	800	&	207,505	&	N	&	*	&	*	&	*	&	*	&	*	&	*	&	*	&	17	&	0.3	\\
{\it brock800\_2}	&	800	&	208,166	&	N	&	*	&	*	&	*	&	*	&	*	&	*	&	*	&	18	&	0.4	\\
{\it brock800\_3}	&	800	&	207,333	&	N	&	*	&	*	&	*	&	*	&	*	&	*	&	*	&	17	&	0.4	\\
{\it brock800\_4}	&	800	&	207,643	&	26	&	*	&	*	&	*	&	*	&	*	&	3455.58	&	*	&	17	&	0.4	\\
{\it c-fat200-1}	&	200	&	1,534	&	12	&	$<$0.01	&	$<$0.01	&	$<$0.01	&	$<$0.01	&	$<$0.01	&	0.02	&	$<$0.01	&	12	&	$<$0.01	\\
{\it c-fat200-2}	&	200	&	3,235	&	24	&	$<$0.01	&	$<$0.01	&	$<$0.01	&	$<$0.01	&	$<$0.01	&	0.02	&	$<$0.01	&	24	&	$<$0.01	\\
{\it c-fat200-5}	&	200	&	8,473	&	58	&	0.6	&	0.33	&	0.01	&	0.03	&	0.03	&	0.03	&	0.93	&	58	&	0.04	\\
{\it c-fat500-1}	&	500	&	4,459	&	14	&	$<$0.01	&	$<$0.01	&	$<$0.01	&	0.02	&	0.02	&	0.05	&	$<$0.01	&	14	&	$<$0.01	\\
{\it c-fat500-2}	&	500	&	9,139	&	26	&	0.02	&	$<$0.01	&	0.01	&	0.02	&	0.02	&	0.04	&	0.01	&	26	&	0.01	\\
{\it c-fat500-5}	&	500	&	23,191	&	64	&	3.07	&	$<$0.01	&	$<$0.01	&	0.03	&	0.03	&	0.05	&	*	&	64	&	0.11	\\
{\it hamming6-2}	&	64	&	1,824	&	32	&	0.68	&	$<$0.01	&	$<$0.01	&	$<$0.01	&	0.01	&	$<$0.01	&	0.33	&	32	&	$<$0.01	\\
{\it hamming6-4}	&	64	&	704	&	4	&	$<$0.01	&	$<$0.01	&	$<$0.01	&	$<$0.01	&	$<$0.01	&	$<$0.01	&	$<$0.01	&	4	&	$<$0.01	\\
{\it hamming8-2}	&	256	&	31,616	&	128	&	*	&	0.01	&	0.01	&	0.04	&	0.04	&	0.04	&	*	&	128	&	0.67	\\
{\it hamming8-4}	&	256	&	20,864	&	16	&	*	&	$<$0.01	&	0.1	&	0.4	&	0.45	&	0.23	&	*	&	16	&	0.03	\\
{\it hamming10-2}	&	1,024	&	518,656	&	512	&	*	&	0.31	&	-	&	0.36	&	0.371	&	0.13	&	*	&	512	&	95.24	\\
{\it johnson8-2-4}	&	28	&	210	&	4	&	$<$0.01	&	$<$0.01	&	$<$0.01	&	$<$0.01	&	$<$0.01	&	$<$0.01	&	$<$0.01	&	4	&	$<$0.01	\\
{\it johnson8-4-4}	&	70	&	1,855	&	14	&	0.19	&	$<$0.01	&	$<$0.01	&	0.01	&	0.02	&	0.01	&	0.23	&	14	&	$<$0.01	\\
{\it johnson16-2-4}	&	120	&	5,460	&	8	&	20.95	&	0.04	&	0.42	&	0.85	&	0.95	&	0.41	&	22.07	&	8	&	$<$0.01	\\
{\it keller4}	&	171	&	9,435	&	11	&	22.19	&	0.15	&	0.02	&	0.21	&	0.33	&	0.12	&	23.35	&	11	&	$<$0.01	\\
{\it keller5}	&	776	&	225,990	&	N	&	*	&	*	&	*	&	*	&	*	&	*	&	*	&	22	&	0.6	\\
{\it keller6}	&	3361	&	4,619,898	&	N	&	*	&	*	&	*	&	*	&	*	&	*	&	*	&	45	&	99.21	\\
{\it MANN\_a9}	&	45	&	918	&	16	&	1.73	&	$<$0.01	&	$<$0.01	&	$<$0.01	&	$<$0.01	&	$<$0.01	&	2.5	&	16	&	$<$0.01	\\
{\it MANN\_a27}	&	378	&	70,551	&	126	&	*	&	*	&	3.3	&	5.75	&	6.03	&	0.97	&	*	&	125	&	1.74	\\
{\it MANN\_a45}	&	1035	&	533,115	&	345	&	*	&	*	&	*	&	4612.29	&	3431.67	&	250.12	&	*	&	341	&	59.96	\\
{\it p\_hat300-1}	&	300	&	10,933	&	8	&	0.14	&	0.01	&	$<$0.01	&	0.07	&	0.08	&	0.06	&	0.14	&	8	&	$<$0.01	\\
{\it p\_hat300-2}	&	300	&	21,928	&	25	&	831.52	&	0.32	&	0.03	&	0.38	&	0.23	&	0.09	&	854.59	&	24	&	0.03	\\
{\it p\_hat300-3}	&	300	&	33,390	&	36	&	*	&	578.58	&	4.31	&	69.91	&	13.53	&	3.2	&	*	&	26	&	$<$0.01	\\
{\it p\_hat500-1}	&	500	&	31,569	&	9	&	2.38	&	0.07	&	0.04	&	0.33	&	0.35	&	0.12	&	2.44	&	9	&	0.02	\\
{\it p\_hat500-2}	&	500	&	62,946	&	36	&	*	&	159.96	&	1.2	&	63.89	&	3.87	&	0.96	&	*	&	34	&	0.14	\\
{\it p\_hat500-3}	&	500	&	93,800	&	50	&	*	&	*	&	324.23	&	*	&	1428.02	&	311.06	&	*	&	39	&	0.27	\\
{\it p\_hat700-1}	&	700	&	60,999	&	11	&	12.7	&	0.12	&	0.13	&	0.96	&	0.92	&	0.23	&	12.73	&	9	&	0.04	\\
{\it p\_hat700-2}	&	700	&	121,728	&	44	&	*	&	*	&	12.28	&	675.72	&	29.36	&	6.76	&	*	&	26	&	0.15	\\
{\it p\_hat1000-1}	&	1,000	&	122,253	&	10	&	97.39	&	1.33	&	0.41	&	1.89	&	2.29	&	0.69	&	98.48	&	10	&	0.11	\\
{\it p\_hat1000-2}	&	1,000	&	244,799	&	46	&	*	&	*	&	406.71	&	*	&	1359.88	&	382.31	&	*	&	33	&	0.57	\\
{\it san200\_0.7\_1}	&	200	&	13,930	&	30	&	*	&	0.99	&	$<$0.01	&	0.05	&	0.47	&	0.1	&	*	&	16	&	0.01	\\
{\it san200\_0.7\_2}	&	200	&	13,930	&	18	&	*	&	0.02	&	$<$0.01	&	0.072	&	0.04	&	0.03	&	*	&	14	&	$<$0.01	\\
{\it san200\_0.9\_2}	&	200	&	17,910	&	60	&	*	&	13.4	&	0.8	&	18.5	&	5.85	&	1.42	&	*	&	34	&	$<$0.01	\\
{\it san200\_0.9\_3}	&	200	&	17,910	&	44	&	*	&	561.64	&	3.16	&	134.67	&	119.27	&	28.02	&	*	&	31	&	$<$0.01	\\
{\it san400\_0.5\_1}	&	400	&	39,900	&	13	&	*	&	$<$0.01	&	0.1	&	0.11	&	0.11	&	0.1	&	*	&	8	&	$<$0.01	\\
{\it san400\_0.7\_1}	&	400	&	55,860	&	40	&	*	&	*	&	0.35	&	1.59	&	2.94	&	0.74	&	*	&	22	&	0.1	\\
{\it san400\_0.7\_2}	&	400	&	55,860	&	30	&	*	&	*	&	0.1	&	12.71	&	19.51	&	4.79	&	*	&	18	&	$<$0.01	\\
{\it san400\_0.7\_3}	&	400	&	55,860	&	22	&	*	&	5.04	&	2.1	&	9.59	&	10	&	2.77	&	*	&	16	&	$<$0.01	\\
{\it san1000}	&	1000	&	250,500	&	15	&	*	&	0.09	&	0.45	&	43.12	&	8.48	&	2.55	&	*	&	10	&	0.5	\\
\bottomrule
\bottomrule
\end{tabular}
\end{table}

%% file: main.bbl
\begin{thebibliography}{10}
\providecommand{\url}[1]{\texttt{#1}}
\providecommand{\urlprefix}{}

\bibitem{heu1}
{D.~Andrade, M.~Resende, and R.~Werneck}, {\em Fast local search for the
  maximum independent set problem}. Journal of Heuristics 18(4),
  pp.~525--547, 2012.

\bibitem{Augustson:1970:AGT:321607.321608}
J.~G.~Augustson and J. Minker, \emph{An analysis of some graph theoretical
  cluster techniques}. Journal of the ACM 17(4), pp. 571--588, 1970.

\bibitem{babel1990branch}
L. Babel and G. Tinhofer, \emph{A branch and bound algorithm for the maximum
  clique problem}. Mathematical Methods of Operations Research 34(3), pp.
  207--217, 1990.

\bibitem{pajek2006}
V. Batagelj and A. Mrvar, \emph{Pajek datasets}, 2006.
  \urlprefix\url{http://vlado.fmf.uni-lj.si/pub/networks/data/}

\bibitem{RePEc:eee:csdana:v:48:y:2005:i:2:p:431-443}
V. Boginski, S. Butenko, and P.M. Pardalos, \emph{Statistical analysis of
  financial networks}. Computational Statistics \& Data Analysis 48(2), pp.
  431--443, 2005.

\bibitem{Bomze99themaximum}
I.~M.~Bomze, M. Budinich, P.~M. Pardalos, and M. Pelillo, \emph{The Maximum
  Clique Problem}. {Handbook of Combinatorial Optimization}, Kluwer
  Academic Publishers, pp. 1--74, 1999.

\bibitem{Bonner:1964:CT:1662386.1662389}
R.~E.~Bonner, \emph{On some clustering techniques}. IBM Journal of Research and Development 8(1),
  pp. 22--32, 1964.

\bibitem{brouwer}
A.~E.~Brouwer, J.~B. Shearer, N.~J.~A. Sloane, and W.~D. Smith, \emph{A new table of
  constant weight codes}. IEEE Transactions on Information Theory 36(6), pp.
  1334--1380, 1990.

\bibitem{pardalos}
R. Carraghan and P. Pardalos, \emph{An exact algorithm for the maximum clique
  problem}. Operations Research Letters 9(6), pp. 375--382, 1990.

\bibitem{Chakrabarti:2006:GML:1132952.1132954}
D. Chakrabarti and C. Faloutsos, \emph{Graph mining: Laws, generators, and
  algorithms}. ACM Computing Surveys 38(1), 2006.

\bibitem{doi:10.1207/s15327906mbr2302_6}
L.~M. Collins and C.~W. Dent.
\newblock {\em Omega: A general formulation of the rand index of cluster recovery
  suitable for non-disjoint solutions}.
\newblock Multivariate Behavioral Research 23(2), pp. 231--242, 1988.

\bibitem{corman2002}
S.~R. Corman, T.~Kuhn, R.~D. Mcphee, and K.~J. Dooley.
\newblock {\em {Studying} complex discursive systems: {Centering} resonance
  analysis of communication}.
\newblock Human Communication Research 28(2), 157--206, 2002.

\bibitem{Davis97theuniversity}
T.~A.~Davis and Y. Hu, \emph{The university of florida sparse matrix
  collection}. ACM Transactions on Mathematical Software (TOMS) 38(1), pp.
  1--25, 2011.

\bibitem{Domingos:2001:MNV:502512.502525}
P. Domingos and M. Richardson, \emph{Mining the network value of customers}, in
  {Proceedings of the 7th ACM SIGKDD (KDD'01)}, ACM, New York, NY, USA, pp. 57--66, 2001.

\bibitem{sea}
D. Eppstein and D. Strash.
\newblock \emph{Listing all maximal cliques in large sparse real-world graphs},
\newblock in {Proceedings of the 10th International Conference on Experimental Algorithms (SEA'11)}, P.~M.~Pardalos and S.~Rebennack (Eds.), Springer-Verlag Berlin Heidelberg, pp. 364--375. 2011.



\bibitem{Faloutsos:1999:PRI:316188.316229}
M. Faloutsos, P. Faloutsos, and C. Faloutsos, \emph{On power-law relationships
  of the internet topology}, in {Proceedings of the Conference on Applications,
  Technologies, Architectures, and Protocols for Computer Communication},
  SIGCOMM '99, ACM, New York, NY, USA, pp. 251--262, 1999.

\bibitem{Ferronato20083922}
M. Ferronato, C. Janna, G. Gambolati, \emph{Mixed constraint preconditioning in computational contact mechanics}.
  Computer Methods in Applied Mechanics and Engineering 197(45-48), pp. 3922--3931, 2008.

\bibitem{Fortunato_2010}
S. Fortunato, \emph{Community detection in graphs}. Physics Reports 486(3),
  pp. 75--174, 2010.

\bibitem{Garey:1979:CIG:578533}
M.~R.~Garey and D.~S.~Johnson, \emph{Computers and intractability: A guide to the theory of NP-Completeness}. W. H. Freeman \& Co., New York, NY, USA, 1979.

\bibitem{2011JSMTE..02..017G}
S.~{Gregory}.
\newblock \emph{Fuzzy overlapping communities in networks}.
\newblock {Journal of Statistical Mechanics: Theory and Experiment} 2011(2), 2011.

\bibitem{heu2}
{A.~Grosso, M.~Locatelli, and W.~Pullan}, {\em Simple ingredients leading
  to very efficient heuristics for the maximum clique problem}. Journal of
  Heuristics 14(6), pp.~587--612, 2008.

\bibitem{Gutin2004}
G. Gutin, J. Gross, J. Yellen, \emph{Discrete Mathematics \& Its Applications}. Handbook of Graph Theory, CRC Press, 2004.

\bibitem{hall2001}
B.~H. Hall, A.~B. Jaffe, and M.~Trajtenberg.
\newblock \emph{{The} {NBER} patent citation data file: {Lessons}, insights and
  methodological tools}.
\newblock {Technical Report, {NBER} {Working} {Paper}}, 8498, 2001.

\bibitem{2011JSMTE..01..023H}
F.~{Havemann}, M.~{Heinz}, A.~{Struck}, and J.~{Gl{\"a}ser}.
\newblock \emph{Identification of overlapping communities and their hierarchy by
  locally calculating community-changing resolution levels}.
\newblock {Journal of Statistical Mechanics: Theory and Experiment} 1(23), 2011.

\bibitem{Horaud:1989:SCT:68871.68875}
R. Horaud and T. Skordas, \emph{Stereo correspondence through feature grouping
  and maximal cliques}. IEEE Transactions on Pattern Analysis and Machine Intellence 11(11), pp.
  1168--1180, 1989.

\bibitem{foldoc}
D. Howe, \emph{Foldoc: Free on-line dictionary of computing}.
  \urlprefix\url{http://foldoc.org/}.

\bibitem{ARI_paper_1985}
L.~Hubert and P.~Arabie.
\newblock \emph{Comparing partitions}.
\newblock {Journal of Classification} 2(1), pp. 193--218, 1985.

\bibitem{dimacs}
D. Johnson and M.A.~Trick, Editors, \emph{Cliques, coloring and satisfiability:
  Second dimacs implementation challenge}. DIMACS Series on Discrete
  Mathematics and Theoretical Computer Science 26, 1996.

\bibitem{hep-th-kdd}
\emph{Kdd cup, 2003}.
  \urlprefix\url{http://www.cs.cornell.edu/projects/kddcup/index.html}.

\bibitem{eatRS}
G.~R.~Kiss, C.~Armstrong, R.~Milroy, and L.~Piper. 
\newblock \emph{An associative thesaurus of English and its computer analysis}.
\newblock {The Computer and Literary Studies}, Edinburgh University Press, 1973.

\bibitem{konc2007improved}
J. Konc and D. Jane\v{z}i\v{c}, \emph{An improved branch and bound algorithm for the
  maximum clique problem}.
  MATCH Communications in Mathematical Computer Chemistry 58, pp. 569--590, 2007.

\bibitem{kumar:extracting}
R. Kumar, P. Raghavan, S. Rajagopalan, and A. Tomkins, \emph{Extracting
  Large-Scale Knowledge Bases from the Web}, in In Proceedings of the 25th International Conference on Very Large Data Bases (VLDB'99), M.~P.~Atkinson, M.~E.~Orlowska, P.~Valduriez, S.~B.~Zdonik, and M.~L.~Brodie (Eds.), Morgan Kaufmann Publishers Inc., San Francisco, CA, USA, pp. 639--650,1999.

\bibitem{2008arXiv0802.1218L}
A.~{Lancichinetti}, S.~{Fortunato}, and J.~{Kertesz}.
\newblock \emph{Detecting the overlapping and hierarchical community structure of
  complex networks}.
\newblock {New Journal of Physics} 11(3), p. 033015, 2009.

\bibitem{2008PhRvE..78d6110L}
A.~{Lancichinetti}, S.~{Fortunato}, and F.~{Radicchi}.
\newblock \emph{Benchmark graphs for testing community detection algorithms}.
\newblock {Physical Review E} 78(4), p. 046110, 2008.

\bibitem{stanford_dataset}
J. Leskovec, \emph{Stanford large network dataset collection}.
  \urlprefix\url{http://snap.stanford.edu/data/index.html}.

\bibitem{web-google}
\emph{Google programming contest, 2002}.
  \urlprefix\url{http://www.google.com/programming-contest/}.

\bibitem{leskovec2007}
J.~Leskovec, L.~Adamic, and B.~Adamic.
\newblock {\em {The} dynamics of viral marketing}.
\newblock ACM Transactions on the Web 1(1), 2007.

\bibitem{leskovec2010}
J.~Leskovec, D.~Huttenlocher, and J.~Kleinberg.
\newblock {\em {Predicting} positive and negative links in online social networks},
\newblock In Proceedings of the 19th International Conference on World Wide Web (WWW'10), pp. 641Ð650, ACM, New York, NY, 2010.

\bibitem{leskovec2007-2}
J.~Leskovec, J.~Kleinberg, and C.~Faloutsos.
\newblock {\em {Graph} evolution: {Densification} and shrinking diameters}.
\newblock ACM Transactions on Knowledge Discovery from Data 1(1), 2007.

\bibitem{Leskovec:2005:GOT:1081870.1081893}
J. Leskovec, J. Kleinberg, and C. Faloutsos, \emph{Graphs over time:
  Densification laws, shrinking diameters and possible explanations}, in
  {Proceedings of the eleventh ACM SIGKDD International Conference on
  Knowledge Discovery in Data Mining (KDD'05)}, ACM,
  New York, NY, USA, pp. 177--187, 2005.

\bibitem{Leydesdorff}
L. Leydesdorff, \emph{On the normalization and visualization of author
  co-citation data: Salton's cosine versus the jaccard index}. Journal of the Association for Information Science and Technology 59(1), pp. 77--85, 2008.

\bibitem{a6040618}
C. McCreesh and P. Prosser.
\newblock \emph{Multi-threading a state-of-the-art maximum clique algorithm}.
\newblock Algorithms 6(4), pp. 618--635, 2013.

\bibitem{Newman06042004}
M.~E.~J.~Newman, \emph{Coauthorship networks and patterns of scientific
  collaboration}, in Proceedings of the National Academy of Sciences of the United
  States of America 101, pp. 5200--5205, 2004.

\bibitem{cliquer}
S. Niskanen and P.R.J. \"{O}sterg{\aa}rd, \emph{Cliquer user's guide, version
  1.0}, Technical Report T48, Communications Laboratory, Helsinki University of
  Technology, Espoo, Finland, 2003.

\bibitem{ostergard}
P.~R.~J. \"{O}sterg{\aa}rd, \emph{A fast algorithm for the maximum clique
  problem}. Discrete Applied Mathematics 120(1-3), pp. 197--207, 2002.

\bibitem{cite-key}
G.~{Palla}, I.~{Der{\'e}nyi}, I.~{Farkas}, and T.~{Vicsek}.
\newblock {Uncovering the overlapping community structure of complex networks
  in nature and society}.
\newblock {\em \nat} 435, pp. 814--818, 2005.

\bibitem{dianapaper}
D.~Palsetia, M.~M.~Patwary, A.~Agrawal, and A.~Choudhary.
\newblock \emph{Excavating social circles via user interests}.
\newblock {Social Network Analysis and Mining} 4(1), pp. 1--12, 2014.

\bibitem{citeulike:4058448}
P. M. Pardalos and J. Xue, \emph{{The maximum clique problem}}. Journal of
  Global Optimization 4, pp. 301--328, 1994.

\bibitem{1211348}
M. Pavan and M. Pelillo, \emph{A new graph-theoretic approach to clustering and
  segmentation}, in {Proceedings of the IEEE Conference on
  Computer Vision and Pattern Recognition (CVPR'03)}, IEEE
  Computer Society, Washington, DC, USA, pp. 145--152, 2003.

\bibitem{prosser2012}
P. Prosser.
\newblock \emph{Exact algorithms for maximum clique: A computational study}.
\newblock {Algorithms}, 5(4), pp. 545--587, 2012.

\bibitem{richardson2003}
M.~Richardson, R.~Agrawal, and P.~Domingos.
\newblock {\em {Trust} management for the semantic web},
\newblock In Proceedings of the Second International Semantic Web Conference (ISWC), Springer-Verlag Berlin Heidelberg, pp. 351--368, 2003.

\bibitem{5586496}
S. Sadi, S. \"{O}\u{g}\"{u}d\"{u}c\"{u}, and A.S. Uyar, \emph{An efficient
  community detection method using parallel clique-finding ants}, in
  {Proceedings of the IEEE Congress on Evolutionary Computation}, pp. 1--7, 2010.

\bibitem{SanSegundo}
P. San~Segundo, D. Rodr\'{\i}guez-Losada, and A. Jim\'{e}nez, \emph{An exact
  bit-parallel algorithm for the maximum clique problem}. Computers \& Operations Research 38(2),
  pp. 571--581, 2011.

\bibitem{citeulike:7905505}
E. Tomita and T. Seki, \emph{{An efficient branch-and-bound algorithm for
  finding a maximum clique}}, in {Proceedings of the 4th International
  Conference on Discrete Mathematics and Theoretical Computer Science},
  Springer-Verlag Berlin Heidelberg, pp. 278--289, 2003.

\bibitem{walcom}
{E.~Tomita, Y.~Sutani, T.~Higashi, S.~Takahashi, and M.~Wakatsuki}, {\em A
  simple and faster branch-and-bound algorithm for finding a maximum clique},
  in Proceedings of the 4th International Workshop on Algorithms and Computation (WALCOM), M.~Rahman and S.~Fujita (Eds.), Springer Berlin Heidelberg, pp.~191--203, 2010.
  
\bibitem{19566964}
T. Matsunaga, C. Yonemori, E. Tomita, and M. Muramatsu, \emph{Clique-based data
  mining for related genes in a biomedical database}. BMC Bioinformatics 10, p. 205, 2010.

\bibitem{Turner88}
J. Turner, \emph{Almost all k-colorable graphs are easy to color}, Journal of
  Algorithms 9, pp. 63--82, 1988.

\bibitem{vanRietbergen199569}
B. van Rietbergen, H. Weinans, R. Huiskes, and A. Odgaard, \emph{A new method
  to determine trabecular bone elastic properties and loading using
  micromechanical finite-element models}. Journal of Biomechanics 28(1),
  pp. 69 -- 81, 1995.

\bibitem{wang2009order}
L. Wang, L. Zhou, J. Lu, and J. Yip, \emph{An order-clique-based approach for
  mining maximal co-locations}. Information Sciences 179(19), pp.
  3370--3382, 2009.
 
\bibitem{Xie:2013:OCD:2501654.2501657}
J.~Xie, S.~Kelley, and B.~K. Szymanski.
\newblock \emph{Overlapping community detection in networks: The state-of-the-art and
  comparative study.}
\newblock {ACM Computing Surveys} 45(4), pp. 43:1--43:35, 2013.


\end{thebibliography}
